\documentclass[draftcls,onecolumn]{IEEEtran}%
\usepackage{amsfonts}
\usepackage{amssymb}
\usepackage{amsmath}
\usepackage{graphicx}%
\newtheorem{lemma}{Lemma}

\newtheorem{proposition}{Proposition}

\begin{document}

\title{Wireless Secrecy in Cellular Systems with Infrastructure--Aided Cooperation}
\author{Petar Popovski,~\IEEEmembership{Member,~IEEE} and Osvaldo
Simeone,~\IEEEmembership{Member,~IEEE} }
\maketitle

\begin{abstract}
In cellular systems, confidentiality of uplink transmission with respect to
eavesdropping terminals can be ensured by creating intentional inteference via
scheduling of concurrent downlink transmissions. In this paper, this basic
idea is explored from an information-theoretic standpoint by focusing on a
two-cell scenario where the involved base stations are connected via a
finite-capacity backbone link. A number of transmission strategies are
considered that aim at improving uplink confidentiality under constraints on
the downlink rate that acts as an interfering signal. The strategies differ
mainly in the way the backbone link is exploited by the cooperating downlink-
to the uplink-operated base stations. Achievable rates are derived for both
the Gaussian (unfaded) and the fading cases, under different assumptions on
the channel state information available at different nodes. Numerical results
are also provided to corroborate the analysis. Overall, the analysis reveals
that a combination of scheduling and base station cooperation is a promising
means to improve transmission confidentiality in cellular systems.

\end{abstract}

\section{Introduction}

The ability to ensure transmission confidentiality is becoming a crucial
requirement of many wireless communications systems due to the increasing role
of on-line transactions and new applications that exchange critical personal
data. In information-theoretic terms, \textit{perfect security (or
confidentiality)} implies the impossibility for a given eavesdropping terminal
to harness any information about the transmitted message from its received
signal~\cite{wiretap}. This condition implies an even stronger guarantee than
traditional cryptography, where security relies on the computational
limitations of the eavesdropper (also referred to as the \textit{wiretap}).

Analysis of transmission strategies that are able to meet the requirement of
perfect security in wireless networks is currently an active area of research.
As a brief and partial review of available literature, we first recall that,
following the basic definitions given in \cite{wiretap} of a \textit{wiretap
channel }(consisting of a single source-destination pair and an eavesdropper),
perfect security in a Gaussian wiretap channel with no fading was studied in
\cite{wiretapgaussian}. More recently, attention has turned to the
investigation of the corresponding fading scenario \cite{barros} \cite{el
gamal}, and to multi-user/ multi-antenna Gaussian models \cite{tekin}%
-\cite{Oggier}. In particular, the Gaussian multiple-access wiretap channel
was studied in \cite{tekin}, the multiple access with confidential messages in
\cite{Liang2}, the broadcast channel in \cite{Liang} (parallel broadcast
channels) and \cite{Liu} (multi-antenna broadcast channels), and the
single-link Gaussian MIMO case in \cite{Khisti} \cite{Oggier}.

In this paper, we focus on secure communications for cellular systems,
motivated by the fact that most of confidential transactions are expected to
be conducted over such networks in the near future. Specifically, a novel
basic approach to ensuring confidentiality is proposed that exploits uplink/
downlink scheduling of transmissions in adjacent cells and cooperation at the
base station (BS) level. In so doing, we follow on the line of research opened
by \cite{ElGamal-RelayEavesdrop}, where it was shown that cooperative
transmission, beside being able to improve throughput or reliability (see,
e.g., \cite{kramer}), can also be instrumental in enhancing the
confidentiality of transmission (for a basic relay network). BS cooperation is
currently being widely investigated as a key enabler for high-data rate
infrastructure networks (see, e.g., \cite{somekh} \cite{jwcc}), and is enabled
by the presence of high-capacity backbone links connecting the BSs. The main
contribution of this work is to show that such technology can also bring
significant gains in terms of secure communications.

The proposed techniques aim at securing uplink transmissions from terminals to
a given BS. The basic idea is to schedule downlink BS transmissions at the
same time as the concurrent uplink transmissions of interest, so as to create
intentional interference on the possible eavesdroppers. Cooperation at the BS
level is then used to convey information about the downlink transmission to
the uplink-operated BS (uplink) over a finite--capacity backbone. This enables
the uplink-operated BS to partially mitigate interference from the BS
transmission. The approach is similar to \cite{negi1} \cite{globecom}
\cite{ElGamal-RelayEavesdrop} \cite{letter}, where artificial noise jams the
reception of the eavesdropper, while using techniques to avoid interference at
the intended receiver. In \cite{negi1} this interference mitigation is
obtained by exploiting the structure and reciprocity of multi-antenna fading
channels, while \cite{globecom} \cite{letter} leverage a infinite-capacity
backbone between receiving and jamming antennas. We propose several new
schemes based on the above mentioned basic idea, each of them using a combined
wireless/backbone transmission. The schemes and the corresponding achievable
rates are investigated and compared via analysis and simulations.
\begin{figure}[ptb]
\centering \includegraphics[width=7.3cm]{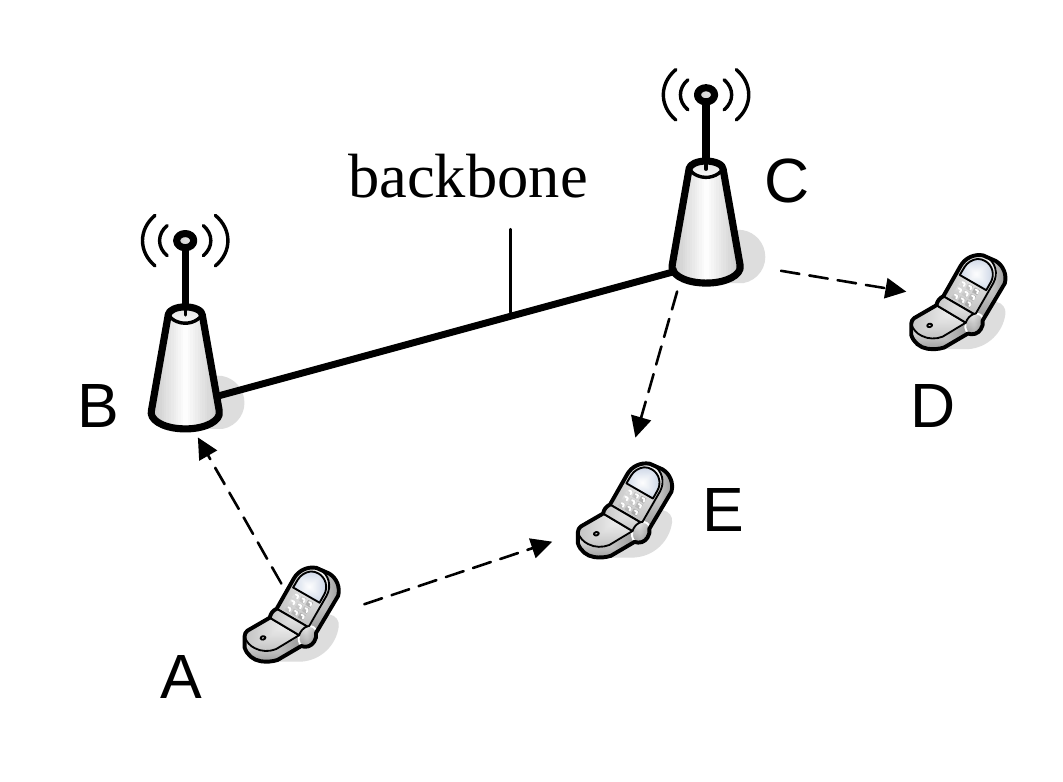} \vspace{-6pt}%
\caption{Illustration of a system with cooperating base stations \textrm{B}
and \textrm{C}$,$ an uplink terminal \textrm{A}, an eavesdropping mobile
station \textrm{E }and a dowlink user \textrm{D}.}%
\label{fig1}%
\end{figure}

\section{System Model and Background}

In this section, we fist introduce the scenario of interest and relevant
quantities, and then investigate the reference case where no infrastructure is
present to enable cooperation between the BSs.

\subsection{Scenario}

We focus on two adjacent cells served by single-antennas BSs as in
Fig.~\ref{fig1} (the contribution of other cells is considered implicitly as
additive noise), where the two BSs are connected by a high capacity, typically
wired, backbone link. The BSs are termed $B$ and $C$, respectively. Terminal
$A$ within the first cell has a message to deliver to $B$ under constraints of
confidentiality with respect to the activity of an eavesdropping terminal $E$.
The eavesdropper is assumed to be within the transmission range of terminal
$A$, as otherwise it would not pose any threat to the confidentiality of $A$'s
message, but also of the adjacent BS $C$. The main idea behind the considered
transmission strategy is that the uplink transmission from $A$ to $B$ can be
scheduled at the same time as the downlink transmission from $C$ towards a
given terminal $D$ in its range. Hence, the transmission from $C$ effectively
acts as a jammer on the reception at $E$. Note this approach is not intended
to secure the communication $C-D$. Also, notice that jamming is thereby
accomplished without exploiting any additional system resource since it is
obtained from a regular downlink transmission.

\subsubsection{System Model}

Formally, terminal $A$ randomly selects a rate-$R_{A}$ message $W_{A}$ from
the set $\{1,...,2^{nR_{A}}\},$ and encodes it via a sequence of $n$ complex
channel inputs $\mathbf{X}_{A}=[X_{A,1}\cdots X_{A,n}]\in\mathbb{C}^{n}$ with
normalized average power constraint \textrm{E}$[|X_{A,i}|^{2}]=P_{A}.$
Encoding takes place through a (possibly stochastic) mapping: $\mathbf{X}_{A}%
$: \{$1,...,2^{nR_{A}}$\}$\rightarrow\mathbb{C}^{n}$
\cite{wiretap} \cite{wiretapgaussian}. Notice that vectors of $n$
symbols are represented throughout the paper by bold letters$.$ At
the same time, BS $C$ transmits a rate-$R_{C}$ downlink message
$W_{C},$ randomly selected from the set $\{1,...,2^{nR_{C}}\}$,
with an average power of $P_{C}$. The actual codebook used by $C$
is assumed to be subject to design and thus depends on the
specific cooperative strategy employed by BSs $B$ and $C.$ This
will be specified for different proposed techniques in the
following sections. The capacity of the backbone link is denoted
by $C_{L}$ and is measured in bit/symbol. We consider bandwidth
that is normalized to $1$ Hz, such that bit/symbol is equivalent
to bit/second (bps). We assume full synchronization between the
transmissions of $A$ and $C$ at the receiver of $B$. Finally, to
account for a worst-case scenario, synchronization is also assumed
at the receiver of eavesdropping terminal $E,$ and the latter is
endowed with information about the codebooks used by $A$\ and $C$.

The complex channel coefficient between any two nodes $U$ and $V$ is denoted
by $h_{UV}$, while the $i-$th symbol transmitted by node $U$ is denoted by
$X_{U,i}$ ($U\in\{A,C\}$ and $V\in\{B,D,E\}$). The signal received by $B$ and
$E$, respectively, at the $i-$th symbol ($i=1,...,n$) reads:
\begin{align}
Y_{B,i}  &  =h_{AB}X_{A,i}+h_{CB}X_{C,i}+N_{B,i}\label{eq:MAC Bob}\\
Y_{E,i}  &  =h_{AE}X_{A,i}+h_{CE}X_{C,i}+N_{E,i} \label{eq:MAC Eve}%
\end{align}
Each noise component $N_{V,i}$ is a complex Gaussian white noise with unit
power, so that if the node $U$ transmits with power $P_{U}$, the corresponding
received signal-to-noise ratio (SNR) at the node $V$ is:
\begin{equation}
\gamma_{UV}=P_{U}|h_{UV}|^{2}. \label{eq:SNRdef}%
\end{equation}
In the most of the paper (Sec. III-VI), we focus on Gaussian (unfaded)
channels, where the channel gains (\ref{eq:SNRdef}) are fixed and
deterministic. In practice, these rates can be achieved, for given channel
realizations, when channel state information is known at the receiver side,
and all the channel gains of interest ($h_{AB},$ $h_{CB},$ $h_{AE},$ $h_{CE})$
are known to terminal $A$, while the channel gains $h_{CB}$ and $h_{CD}$ are
known to the downlink-operated BS $C$. In Sec. VII, the analysis will be
extended to a fading scenario under different assumptions on the channel state
information at the transmitters' side.

Finally, the BS $B$ decodes through a mapping $g(\mathbf{Y}_{B})$:
$\mathbb{C}^{n}\rightarrow\{1,...,2^{nR_{A}}\}.$ According to standard
definitions \cite{wiretap} \cite{wiretapgaussian}, a rate $R_{A}=R_{A,s}$ is
said to be \textit{achievable with perfect secrecy} with respect to
eavesdropper $E$ if, as the number of samples per coding block $n\rightarrow
\infty$: (\textit{a}) the decoding error at BS $B$\ vanishes:%
\begin{equation}
P_{e}=P[g(\mathbf{Y}_{B})\neq W_{A}]\rightarrow0;
\end{equation}
(\textit{b}) the uncertainty (equivocation) $\Delta$ of eavesdropper $E$
regarding $A$'s message, measured as the conditional entropy of $W_{A}$ given
the signal received by $E$ normalized over the unconditional entropy,
satisfies:
\begin{equation}
\Delta=\frac{H(W_{A}|\mathbf{Y}_{E})}{H(W_{A})}\rightarrow1.
\end{equation}

\subsection{Some Useful Functions}

To simplify the presentation of the results in this paper, it is useful to
define the following two functions. The first function $\mathcal{C}%
(\gamma_{UV})$ is the standard capacity of a Gaussian single link with source
$U$ and receiver $V,$ and SNR equal to $\gamma_{UV}$:
\begin{equation}
\mathcal{C}(\gamma_{UV})=\log(1+\gamma_{UV}). \label{eq:CapGaussChannel}%
\end{equation}
The second function $S_{U_{1}V}(R_{U_{2}})$ pertains to the performance of a
multiple--access channel (MAC) with two users $U_{1}$ and $U_{2}$ and receiver
$V.$ It measures the supremum of the achievable rates from $U_{1}$ to $V$ for
a given transmission rate $R_{U_{2}}$ of $U_{2}.$ Notice that rate $R_{U_{2}}$
is not restricted to be within the MAC\ capacity region, that is, it is not
necessarily decodable by $V.$ Given the SNRs $\gamma_{U_{1}V}$ and
$\gamma_{U_{2}V}$, the function is given by:{\small
\begin{equation}
S_{U_{1}V}(R_{U_{2}})=\left\{
\begin{array}
[c]{ll}%
\mathcal{C}(\gamma_{U_{1}V}) & \text{ if }R_{U_{2}}\leq\mathcal{C}\left(
\frac{\gamma_{U_{2}V}}{1+\gamma_{U_{1}V}}\right) \\
{\mathcal{C}(\gamma_{U_{1}V}+\gamma_{U_{2}V})-R_{U_{2}}} & \text{ if
}{\mathcal{C}\left(  \frac{\gamma_{U_{2}V}}{1+\gamma_{U_{1}V}}\right)
<R_{U_{2}}\leq\mathcal{C}(\gamma_{U_{2}V})}\\
\mathcal{C}\left(  \frac{\gamma_{_{U_{1}V}}}{1+\gamma_{U_{2}V}}\right)  &
\text{ if }R_{U_{2}}>\mathcal{C}(\gamma_{U_{2}V})
\end{array}
\right.  . \label{eq:ExampleS_U1V}%
\end{equation}
}

\subsection{Perfect Secrecy Without Backbone Link ($C_{L}=0$)}

\label{sec:C_L=0 scheme}

Here, we briefly discuss the baseline scenario where no backbone link exists
between BSs $B$ and $C$ ($C_{L}=0$)$.$ In such a case, no cooperation via the
backbone link is possible, and we assume that the BS $C$ transmits with a
standard Gaussian codebook $\mathbf{X}_{C}(W_{C})=[X_{C,1}(W_{C})\cdots
X_{C,n}(W_{C})]\in\mathbb{C}^{n},$ where variables $X_{C,i}$ are generated as
complex Gaussian independent with zero mean and power $P_{C}.$ As explained
above, this codebook conveys information to a downlink user $D.$ Given this
set-up, it can be readily seen that the considered approach coincides with the
strategy considered in~\cite{ElGamal-RelayEavesdrop} under the name
\emph{Noise--Forwarding (NF)}. It was shown therein that the secrecy capacity
can be found by considering the compound multiple access channel (MAC), with
two receivers $B$ and $E$ and two transmitters $A$ and $C$. In particular, for
the Gaussian case of interest here, and using the
function~(\ref{eq:ExampleS_U1V}), the result of~\cite{ElGamal-RelayEavesdrop}
(Theorem 3) can be restated as follows.

\begin{proposition}
If BS $C$ transmits in downlink with rate $R_{C}$ and there is no backbone
link ($C_{L}=0$)$,$ the rate $R_{A,s}(R_{C})$ is achievable with perfect
secrecy with respect to eavesdropper $E$\footnote{We define $(x)^{+}=x$ if
$x>0$ and $(x)^{+}=0$ otherwise.}:%
\begin{equation}
R_{A,s}(R_{C})=\left(  S_{AB}(R_{C})-S_{AE}(R_{C})\right)  ^{+},
\label{eq:SecrecyNFstrategy}%
\end{equation}
with $S_{AB}(R_{C})$ and $S_{AE}(R_{C})$ defined in (\ref{eq:ExampleS_U1V}).
\end{proposition}

From (\ref{eq:SecrecyNFstrategy}) it can be seen that an increase in the
secrecy rate can be obtained by either increasing the achievable rate
$S_{AB}(R_{C})$ to the intended destination $B$ or hampering reception of the
eavesdropper (decreasing $S_{AE}(R_{C})).$

\section{Perfect Secrecy with Large-Capacity Backhaul Link ($C_{L}\geq R_{C}%
$)\label{sec: large backbone}}

We now turn to the interesting case where the backhaul link has a capacity
larger than the downlink rate, $C_{L}\geq R_{C}.$ As in the baseline case
considered above, we assume that BS $C$ transmits codewords from a given
rate-$R_{C}$ randomly generated Gaussian codebook. Since $C_{L}\geq R_{C}$, BS
$C$ can communicate the current codeword $\mathbf{X}_{C}(W_{C})$ to the
adjacent BS $B$ by using the backbone link. Therefore, BS $B$ can effectively
cancel the interference signal $\mathbf{X}_{C}(W_{C})$ from the received
signal (\ref{eq:MAC Bob}), leading to the equivalent received signal
\begin{equation}
Y_{B,i}=h_{AB}X_{A,i}+N_{B,i}. \label{Y_B_i}%
\end{equation}
This implies that for any $R_{C}$ we have:
\begin{equation}
S_{AB}(R_{C})=S_{AB}(0)=\mathcal{C}(\gamma_{AB}), \label{eq:}%
\end{equation}
from which the following proposition easily follows.

\begin{proposition}
\label{prop:InfiniteCL} If BS $C$ transmits in downlink with rate $R_{C}$ and
$C_{L}\geq R_{C},$ the rate $R_{A,s}(R_{C})$ is achievable with perfect
secrecy:
\begin{equation}
R_{A,s}(R_{C})=\left(  \mathcal{C}(\gamma_{AB})-S_{AE}(R_{C})\right)  ^{+}
\label{eq:Cinf}%
\end{equation}
with $S_{AE}(R_{C})$ defined in (\ref{eq:ExampleS_U1V}).
\end{proposition}

\textit{Proof}: Follows directly from Theorem 3 of
\cite{ElGamal-RelayEavesdrop} (see discussion in Sec. \ref{sec:C_L=0 scheme}).

\begin{figure}[ptb]
\centering
\includegraphics[width=8.3cm]{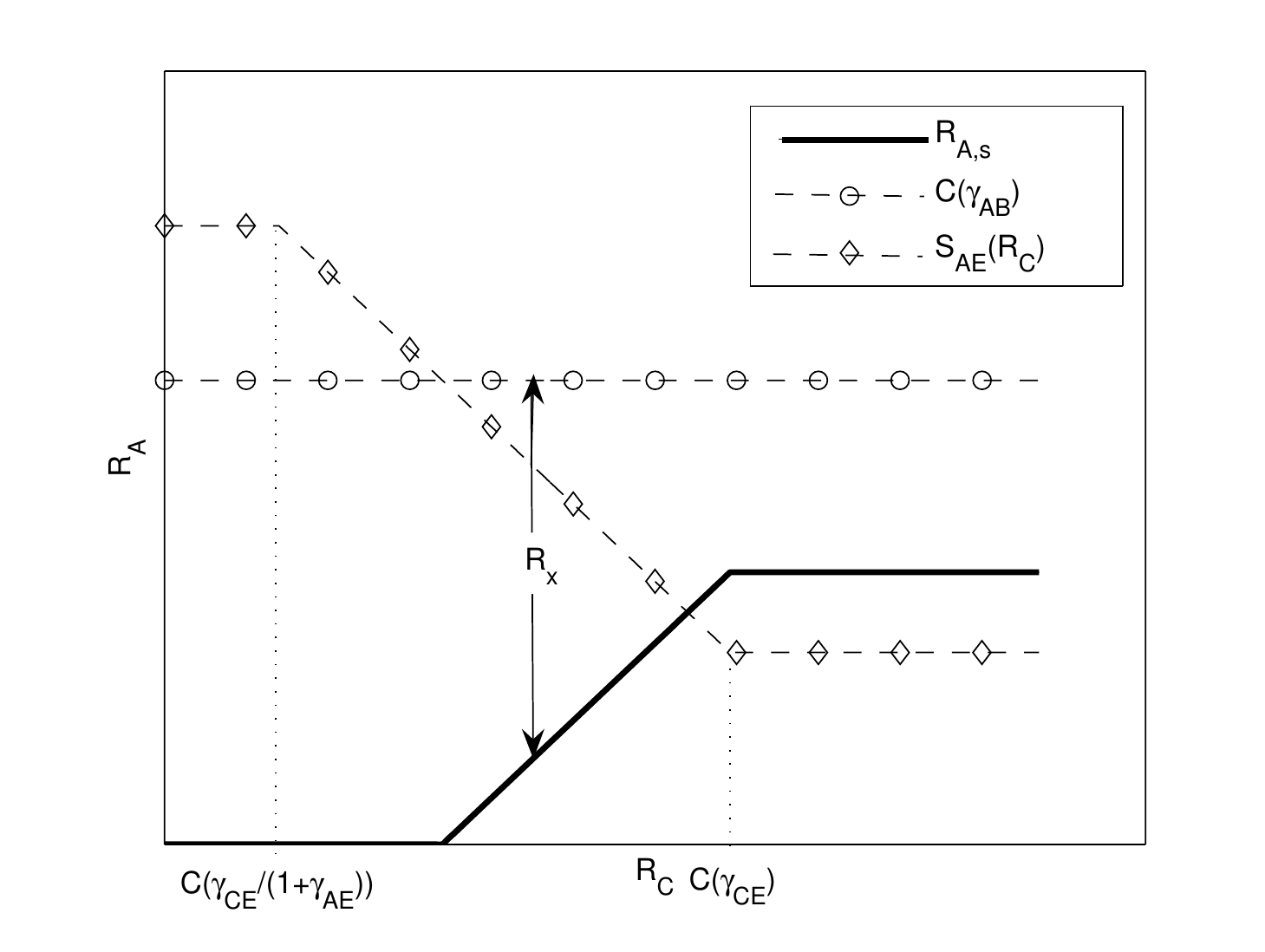}
\caption{The
achievable secrecy rate $R_{A,s}$ in Proposition~\ref{prop:InfiniteCL}. Here
$R_{x}=\mathcal{C}(\gamma_{AB})-R_{A,s}$ is the amount of information spent by
$A$ to \textquotedblleft confound\textquotedblright\ the eavesdropper
\textrm{E}$\ $in order to achieve a rate $R_{A,s}$ with perfect secrecy.}%
\label{CLinfty}%
\end{figure}

The rate (\ref{eq:Cinf}) is plotted in Fig.~\ref{CLinfty} along with the
capacity of the direct link $\mathcal{C}(\gamma_{AB})$ and the maximum
achievable rate at the eavesdropper $S_{AE}(R_{C})$ for $\gamma_{AB}%
=7,\gamma_{AE}=15,\gamma_{CE}=10$. A relevant quantity that can be observed
from the figure is the rate $R_{x}=\mathcal{C}(\gamma_{AB})-R_{A,s}.$ This can
be interpreted as the \textit{rate loss }that terminal $A$ must sacrifice to
the aim of \textquotedblleft confounding\textquotedblright\ the eavesdropper
$E$ and thus achieving rate $R_{A,s}$ with perfect secrecy. For this
particular example, when $R_{C}=0$, the single--user link $A-E$ is less noisy
than the link $A-B$ and therefore the secrecy capacity is zero. As the
downlink rate $R_{C}$ increases, while the achievable rate $\mathcal{C}%
(\gamma_{AB})$\ on the link $A-B$ is clearly unaffected (see (\ref{Y_B_i})),
the rate decodable by the eavesdropper $S_{AE}(R_{C})$ decreases (for $R_{C}$
large enough), and thus a positive secrecy rate is obtained as soon as
$S_{AE}(R_{C})<\mathcal{C}(\gamma_{AB})$. In particular, the secrecy rate
$R_{A,s}$ increases linearly with $R_{C}$ until it reaches the maximum value
$\left(  \mathcal{C}(\gamma_{AB})-\mathcal{C}\left(  \frac{\gamma_{AE}%
}{1+\gamma_{CE}}\right)  \right)  ^{+}$ for
$R_{C}\geq\mathcal{C}(\gamma _{CE})$. It can be easily seen that
this value of $R_{A,s}$ corresponds to the case where the signal
from $C$ acts as a Gaussian noise with power $\gamma_{CE},$ which
is known to be worst--case jammer on~$E$ (see, e.g.,
\cite{Diggavi}).

Finally, two remarks on the case at hand of large-capacity backbone link
($C_{L}\geq R_{C}$) are in order, that will be compared in the next sections
with the complementary case where $C_{L}<R_{C}$. (\textit{a}) With a
large-capacity backbone, the secrecy rate $R_{A,s}$ is a non-decreasing
function of the downlink rate $R_{C}.$ (\textit{b}) With a large-capacity
backbone, the value of the inter-BS channel gain $\gamma_{CB}$ is irrelevant
to the system performance. This clearly contrasts with the case of $C_{L}=0$
studied in Sec. \ref{sec:C_L=0 scheme}: for instance, for the chosen SNRs in
the example on Fig.~\ref{CLinfty}, if in addition we assume $\gamma_{CB}%
\leq\gamma_{CE}$, it can be seen from~(\ref{eq:SecrecyNFstrategy})
that with $C_{L}=0$ the secrecy rate $R_{A,s}$ is identically
zero.

\section{Quantization--based Transmission Strategies}

In this section, we start the analysis of the secrecy capacity for the case
where the backbone link capacity is smaller than the downlink rate,
$C_{L}<R_{C}.$ In such as case, different strategies can be devised by the BS
$C$ in order to provide the adjacent BS $B$ with some information about the
downlink transmitted waveform $\mathbf{X}_{C}$ in order to enable interference
mitigation at $B$ and thus improve the secrecy rate $R_{A,s}$ (recall the
discussion about (\ref{eq:SecrecyNFstrategy})). In this section, we describe
two strategies based on source coding arguments (quantization) for
transferring information from $C$ to $B$, while the next section proposes
strategies based on channel coding principles. For the strategies considered
in this section, we assume, as above, that the BS $C$ employs a standard
randomly generated Gaussian codebook.

\subsection{Elementary Quantization}

The first considered approach is based on quantizing the downlink codeword
$\mathbf{X}_{C}(W_{C})$ via a rate-$C_{L}$ Gaussian codebook.
Quantization/compression is done by using standard joint typicality-based
vector quantization~\cite{cover} and does not exploit here any side
information available at the receiver (\emph{elementary quantization}). Given
its optimality in a rate-distortion sense, here we consider a Gaussian test
channel, which we represent for convenience in the forward
form~\cite{gallager}:
\begin{equation}
\hat{X}_{C,i}=X_{C,i}+Q_{i}, \label{eq:EQtestchannel}%
\end{equation}
where $Q_{i}$ is i.i.d. complex Gaussian quantization noise with power
$\sigma_{Q}^{2}$. From basic rate-distortion theory, it follows that the
following condition should be satisfied:
\begin{equation}
I(X_{C};\hat{X}_{C})=C_{L} \label{eq:EQcondition}%
\end{equation}
which, as mentioned above, reflects the fact that the quantization process at
$C$ is oblivious to the fact that there is a parallel wireless link between
$C$ and $B$ that conveys side information. The quantization error power
$\sigma_{Q}^{2}$ can then be found from (\ref{eq:EQcondition}) as $\sigma
_{Q}^{2}=P_{C}/(2^{C_{L}}-1)$ (since $I(X_{C};\hat{X}_{C})=\log_{2}\left(
1+P_{C}/\sigma_{Q}^{2}\right)  ),$ so that the SNR\ on the equivalent channel
(\ref{eq:EQtestchannel}) reads
\begin{equation}
\gamma_{Q}=\frac{P_{C}}{\sigma_{q}^{2}}=2^{C_{L}}-1. \label{gamma Q}%
\end{equation}
It is remarked that the quantization codebook is assumed to be known to the BS
$B$, which uses the received index from the backbone link to decompress the
signal into $\mathbf{\hat{X}}_{C}$. The following proposition provides the
rate achievable with this strategy (see proof in Appendix-A).

\begin{proposition}
\label{prop:quant} If BS $C$ transmits in downlink with rate $R_{C}$, the
elementary quantization-based strategy achieves with perfect secrecy the rate
$R_{A,s}(R_{C})$ given by:%
\begin{equation}
R_{A,s}(R_{C})=\left(  S_{AB}^{EQ}(R_{C})-S_{AE}(R_{C})\right)  ^{+}
\label{eq: quantization}%
\end{equation}
with $S_{AE}(R_{C})$ defined in (\ref{eq:ExampleS_U1V}),
\begin{equation}
S_{AB}^{EQ}(R_{C})=\left\{
\begin{array}
[c]{l}%
\mathcal{C}(\gamma_{AB})\qquad\text{ }\text{if }R_{C}\leq\mathcal{C}\left(
\frac{\gamma_{CB}}{1+\gamma_{AB}}+\gamma_{Q}\right) \\
\mathcal{C}_{sum}-R_{C}\quad\text{if }\mathcal{C}\left(  \frac{\gamma_{CB}%
}{1+\gamma_{AB}}+\gamma_{Q}\right)  <R_{C}\leq\mathcal{C}(\gamma_{CB}%
+\gamma_{Q})\\
\mathcal{C}_{sum}-\mathcal{C}(\gamma_{CB}+\gamma_{Q})\qquad\text{ }\text{if
}R_{C}>\mathcal{C}(\gamma_{CB}+\gamma_{Q})
\end{array}
\right.  \label{eq:Sq}%
\end{equation}
and $\mathcal{C}_{sum}=\log_{2}\left(
2^{C_{L}}(1+\gamma_{AB})+\gamma _{CB}\right)  .$
\end{proposition}

It can seen that, unlike the large-backbone case of Proposition
\ref{prop:InfiniteCL}, here the achievable rate (\ref{eq:
quantization}) is not a monotonically increasing function of
$R_{C}$ since the latter affects (decreases) also
$S_{AB}^{EQ}(R_{C}).$ Moreover, it can be shown that only for
$C_{L}\rightarrow\infty$, the rate (\ref{eq: quantization}) tends
to the large-backbone secrecy rate (\ref{eq:Cinf}) due to the
residual quantization noise for any finite $C_{L}$. In practice
(and in our evaluations in Sections~\ref{sec: numerical}
and~\ref{sec:ExtensionFading}, whenever the instantaneous rate
$R_C \geq C_L$, then we do not use quantization, but transfer the
message completely over the backhaul.

\subsection{Wyner--Ziv Quantization}

The approach presented above can be improved by designing the quantization
scheme according to Wyner-Ziv compression with side information at the decoder
\cite{Wyner-Ziv} (\emph{Wyner-Ziv quantization}). In fact, the wireless signal
$Y_{B}$ received by BS $B$ is correlated with the signal $X_{C}$ transmitted
by BS $C$ and can thus be used as side information at the decoder$.$ From
\cite{Wyner-Ziv}, the following relationship should now hold:
\begin{equation}
I(X_{C};\hat{X}_{C}|Y_{B})=C_{L}. \label{eq:WZcondition}%
\end{equation}
By using the Gaussian forward test channel~(\ref{eq:EQtestchannel}), the power
of the quantization noise $\sigma_{Q}^{2}$ and the respective equivalent
quantization SNR $\gamma_{Q}$ can be easily derived (see Appendix-B) leading
to the equivalent SNR
\begin{equation}
\gamma_{Q}=\frac{P_{C}}{\sigma_{q}^{2}}=(2^{C_{L}}-1)\left(  1+\frac
{\gamma_{CB}}{1+\gamma_{AB}}\right)  . \label{eq:WZgammaq}%
\end{equation}
Achievable rates with Wyner-Ziv compression then follow directly from
Proposition \ref{prop:quant} by simply replacing $\gamma_{Q}$ in
(\ref{gamma Q}) with (\ref{eq:WZgammaq}). It is noted that if $\gamma_{CB}=0,$
this scheme has clearly no advantage over the elementary quantization
considered above due to the absence of useful side information at the receiver.

\section{Superposition Coding-based Transmission Strategies}

\label{sec: superposition}

Here we investigate a channel coding-based strategy to exploit the backbone
link with capacity satisfying the condition $C_{L}<R_{C}$. The strategy is
based on rate--splitting encoding at BS $C$ so that, differently from the
previous sections, here $C$ changes the format of its wireless transmission to
facilitate the transfer of information over the backbone. It is also noted
that this assumption requires downlink terminal $D$ to modify its decoding
strategy accordingly (see details below). The message $W_{C}$ is transmitted
by sending two independent messages $W_{C1},$ $W_{C2}$ with rates $R_{C1},$
$R_{C2}$, respectively such that:
\begin{equation}
R_{C}=R_{C1}+R_{C2}, \label{eq:SumRatesR_C}%
\end{equation}
where $R_{C}$ is determined by the capacity of the downlink transmission by
$C$:
\begin{equation}
R_{C}=\mathcal{C}(\gamma_{CD}). \label{eq:SUPtotalrate}%
\end{equation}
The two messages are combined by using superposition coding, such that the
$i$th symbol sent by $C$ is:
\begin{equation}
X_{C,i}=\sqrt{\alpha}X_{C1,i}+\sqrt{1-\alpha}X_{C2,i}, \label{eq:SupCodingC}%
\end{equation}
where $\alpha$ is the power--division coefficient and $0\leq\alpha\leq1$.
Notice that, unlike the previously described quantization-based scheme, here
the downlink channel gain $\gamma_{CD}$ plays an important role, since any
modification in the design of the transmission scheme at BS $C$ (i.e., rates
($R_{C1},R_{C2}$) and coefficient $\alpha$) has to guarantee successful
decoding at terminal $D$. To elaborate, we assume that decoding at $D$ is
carried out via successive interference cancellation, such that $W_{C1}$ is
first decoded and subtracted and then $W_{C2}$ is decoded. Such a decoding
imposes the following conditions on rates ($R_{C1},R_{C2}$) and coefficient
$\alpha$:
\begin{align}
R_{C1}  &  =\log_{2}\left(  1+\frac{\alpha\gamma_{CD}}{1+(1-\alpha)\gamma
_{CD}}\right) \label{eq:RateC1}\\
R_{C2}  &  =\log_{2}(1+(1-\alpha)\gamma_{CD}). \label{eq:RateC2}%
\end{align}
It can be easily seen that for any $0\leq\alpha\leq1,$ condition
(\ref{eq:SUPtotalrate}) is satisfied ($R_{C1}+R_{C2}=\mathcal{C}(\gamma_{CD}%
)$) and we have freedom to chose $\alpha$.

The basic idea of this strategy is to send one of the messages, either
$W_{C1}$ or $W_{C2}$ (i.e., either the one decoded first or last by downlink
user $D$), over the backbone. This implies either $R_{C1}=C_{L}$ or
$R_{C2}=C_{L},$ respectively. It is noted that once either of the latter
condition is specified, this choice, by way of (\ref{eq:RateC1}%
)-(\ref{eq:RateC2}), uniquely determines the value of $\alpha$ and, from
(\ref{eq:SumRatesR_C}), the remaining rate. As we will see in the next
sections, the choice of which message to send over the backbone drastically
impact the achievable secrecy rate, and neither strategy dominates the other.

\subsection{Sending the Message Decoded Last by \textrm{D} ($W_{C2}$)
\label{Sec:Wc2}}

In this first case, we set $R_{C2}=C_{L}$, which, from (\ref{eq:RateC2}),
determines the following value of $\alpha$:
\begin{equation}
\alpha=\alpha_{2}=1-\frac{2^{C_{L}}-1}{\gamma_{CD}},
\label{eq:XC1 over backhaul}%
\end{equation}
and the rate $R_{C1}=R_{C}-C_{L}$. BS $B$ can then uses $W_{C2}$ to cancel
$\mathbf{X}_{C2}(W_{C2})$ from its wireless received signal $\mathbf{Y}_{B}$,
such that the resulting received wireless signal at $B$ at the instance $i$ is
given by:
\begin{equation}
Y_{B,i}=h_{AB}X_{A,i}+h_{CB}\sqrt{\alpha_{2}}X_{C1,i}+N_{B,i}.
\label{eq:MAatBafterXC1cancel}%
\end{equation}
The following lemma follows (see proof in Appendix-C).

\begin{lemma}
\label{prop: SUP1} If BS $C$ transmits in downlink with rate $R_{C}$ using the
superposition coding scheme with (\ref{eq:XC1 over backhaul}), the maximum
rate achievable on the link $A$-$B$ is given by:%
\begin{equation}
S_{AB}^{(\alpha_{2})}(R_{C})=\left\{
\begin{array}
[c]{ll}%
\min\{\mathcal{C}(\gamma_{AB}),\mathcal{C}(\gamma_{AB}+\alpha_{2}\gamma
_{CB})-(R_{C}-C_{L})\} & \text{if }R_{C}-C_{L}<\mathcal{C}%
(\alpha_{2}\gamma_{CB})\\
\mathcal{C}\left(
\frac{\gamma_{AB}}{1+\alpha_{2}\gamma_{CB}}\right) &
\text{otherwise }%R_{C}-C_{L}\geq\mathcal{C}(\alpha_{2}\gamma_{CB})
\end{array}
\right.  \label{eq: S_AB2}%
\end{equation}

\end{lemma}

\subsection{Sending the Message Decoded First by \textrm{D} ($W_{C1}$)
\label{Sec:Wc1}}

When $W_{C1}$ is sent over the backbone, we set $R_{C1}=C_{L}$, resulting in
\begin{equation}
\alpha=\alpha_{1}=\frac{1-2^{-C_{L}}}{1-1/(1+\gamma_{CD})}
\label{eq:XC2 over backhaul}%
\end{equation}
and $R_{C2}=R_{C}-C_{L}.$ After cancelling out $\mathbf{X}_{C1}(W_{C1})$, the
multiple access channel at $B$ is given as:
\begin{equation}
Y_{B,i}=h_{AB}X_{A,i}+h_{CB}(1-\sqrt{\alpha_{1}})X_{C2,i}+N_{B,i}.
\label{eq:MAatBafterXC1cancel}%
\end{equation}
The following result follows from the same arguments as in Appendix-C.

\begin{lemma}
\label{prop: SUP2} If BS $C$ transmits in downlink with rate $R_{C}$ using the
superposition coding scheme with (\ref{eq:XC2 over backhaul}), the maximum
rate achievable on the link $A$-$B$ is given by:%
\begin{equation}
S_{AB}^{(\alpha_{1})}(R_{C})=\left\{
\begin{array}
[c]{ll}%
\mathrm{min}\{\mathcal{C}(\gamma_{AB}),\mathcal{C}(\gamma_{AB}+(1-\alpha
_{1})\gamma_{CB})-(R_{C}-C_{L})\} & \text{if }R_{C}-C_L<\mathcal{C}%
((1-\alpha_{1})\gamma_{CB})\\
\mathcal{C}\left(
\frac{\gamma_{AB}}{1+(1-\alpha_{1})\gamma_{CB}}\right) &
\text{otherwise }%R_{C}\geq\mathcal{C}((1-\alpha_{1})\gamma_{CB})
\end{array}
\right.  \label{eq: S_AB1}%
\end{equation}

\end{lemma}

\subsection{Achievable Secrecy Rate with Superposition
Coding\label{sec: achievable SUP}}

Accounting for both options of sending either $W_{C1}$ or $W_{C2}$ over the
backbone, we can now state the following result.

\begin{proposition}
If BS $C$ transmits in downlink with rate $R_{C},$ the superposition-based
strategy achieves the following rate $R_{A,s}(R_{C})$ with perfect
secrecy:{\small
\begin{equation}
R_{A,s}(R_{C})=\left(  S_{AB}^{SUP}(R_{C})-S_{AE}(R_{C})\right)  ^{+}
\label{eq: SUPmaxachievable}%
\end{equation}
with}%
\begin{equation}
S_{AB}^{SUP}(R_{C})=\max_{i=1,2}\left\{  S_{AB}^{(\alpha_{i})}(R_{C})\right\}
, \label{eq: S SUP}%
\end{equation}
{\small where }$S_{AB}^{(\alpha_{i})}(R_{C})$ are defined in (\ref{eq: S_AB2})
and (\ref{eq: S_AB1}), and $S_{AE}(R_{C})$ is given by (\ref{eq:ExampleS_U1V}).
\end{proposition}

The proposition follows from Lemmas \ref{prop: SUP2}\ and \ref{prop: SUP1} and
similar arguments as in the proof of Proposition 1
\cite{ElGamal-RelayEavesdrop}. In particular, following such arguments, one
should calculate the maximum rate decodable by the eavesdropper $E$ for given
$R_{C}$ and for the rate splitting strategy. It can be shown that this maximum
rate is indeed $S_{AE}(R_{C})$ as in (\ref{eq: SUPmaxachievable})$,$ that is,
it is the same rate that we would have if BS $C$ had used a single-rate
Gaussian codebook. This is because from~(\ref{eq:SumRatesR_C}%
),(\ref{eq:RateC1}), and~(\ref{eq:RateC2}), it can be proved that any of the
superposed messages is decodable if and only if the other is.

A final remark concerns a comparison between the superposition strategy and
elementary quantization. It can be shown by comparing
(\ref{eq: SUPmaxachievable}) and (\ref{eq:Sq}) (with (\ref{gamma Q})) that for
downlink rate $R_{C}\rightarrow\infty$ the performance of both scheme coincide
since $S_{AB}^{SUP}(R_{C})\rightarrow S_{AB}^{EQ}(R_{C})=\mathcal{C}%
_{sum}-\mathcal{C}(\gamma_{CB}+\gamma_{Q}).$

\subsection{Some Comments on the Superposition Strategy\label{sec: comments}}

The achievable secrecy rate (\ref{eq: SUPmaxachievable}) contains
a maximization over the choice of which message should be sent
over the backhaul link. This choice is made so as to optimize the
maximum achievable rate on the link $A$-$B$ (\ref{eq: S SUP}). In
this regard, some general conclusion can be drawn by noticing that
from the assumption $R_{C}=\mathcal{C}(\gamma_{CD}) \geq C_{L}$ it
can be verified that
\begin{equation}
\alpha_{2} \geq 1-\alpha_{1}. \label{eq:alpha2 > 1-alpha1-1}%
\end{equation}
Then the following observations can be made:
\begin{itemize}
    \item For \textit{large} downlink rates $R_{C}$, such that:
    \begin{equation}\label{eq:SupLargeRC_cond}
        R_{C}\geq
        C_{L}+\mathcal{C}(\alpha_{2}\gamma_{CB})=\mathcal{C}(2^{C_{L}}\gamma_{CB})
        \stackrel{(a)}{\geq}C_{L}+\mathcal{C}((1-\alpha_{1})\gamma_{CB})
    \end{equation}
    where (a) follows from~(\ref{eq:alpha2 > 1-alpha1-1})it follows from Lemmas~\ref{prop: SUP1} and~\ref{prop: SUP2}
    that
    \begin{equation}\label{eq:SupLargeRC_rates}
        S_{AB}^{(\alpha_{2})}(R_{C})=\mathcal{C}\left(\frac{\gamma_{AB}}{1+\alpha_{2}\gamma_{CB}}\right)
        \leq
        \mathcal{C}\left(\frac{\gamma_{AB}}{1+(1-\alpha_1)\gamma_{CB}}\right)=S_{AB}^{(\alpha_{1})}(R_{C})
    \end{equation}
    which means that sending $X_{C1}$ over the backbone offers higher achievable rates
    $R_{A}$.
    \item For \textit{low} downlink rates $R_{C}$, such that:
    \begin{equation}\label{eq:SupLowRC_cond}
        R_{C} \leq
        C_{L}+\mathcal{C}((1-\alpha_{1})\gamma_{CB})=
        \leq C_{L}+\mathcal{C}(\alpha_{2}\gamma_{CB})
    \end{equation}
    it follows from Lemmas~\ref{prop: SUP1} and~\ref{prop: SUP2}
    that
    \begin{eqnarray}\label{eq:SupLowRC_rates}
        S_{AB}^{(\alpha_{2})}(R_{C})&=&\min\{\mathcal{C}(\gamma_{AB}),\mathcal{C}(\gamma_{AB}+\alpha_{2}\gamma_{CB})-(R_{C}-C_{L})\} \nonumber \\
        S_{AB}^{(\alpha_{1})}(R_{C}) &=&
        \min\{\mathcal{C}(\gamma_{AB}),\mathcal{C}(\gamma_{AB}+(1-\alpha_1)\gamma_{CB})-(R_{C}-C_{L})\}\nonumber
        \\
        \textrm{thus } \quad S_{AB}^{(\alpha_{2})}(R_{C}) &\geq& S_{AB}^{(\alpha_{1})}(R_{C})
    \end{eqnarray}
    which means that sending $X_{C2}$ over the backbone offers higher achievable rates
    $R_{A}$.
\end{itemize}

We give an intuitive explanation of the previous result. Note that
if a signal contains two superposed messages and one of those
messages is known a priori, then this is equivalent to cancelling
power from the composite message. For example, if
in~(\ref{eq:SupCodingC}) the message $W_{C1}$ is known, then we
cancel the signal $\sqrt{\alpha} X_{C1}$, which corresponds to the
power of $\alpha P_C$. Now we can ask the following question: If
we fix the condition $R_{C1}+R_{C2}=R_C$ and we set one of the
rates ($R_{C1}$ or $R_{C2}$) to be equal to $C_L$, then in which
case we can cancel the maximal amount of power from the composite
message? From the previous discussions, if $R_{Cj}=C_L$ then we
determine $\alpha=\alpha_j$. If $W_{C2}$ is sent over the backhaul
then $j=2$ and the amount of power cancelled is $(1-\alpha_2)
P_C$. If $j=1$, the amount of power cancelled is $\alpha_1 P_C$.
Hence, using the condition~(\ref{eq:alpha2 > 1-alpha1-1}), we
conclude that sending $W_{C2}$ over the backhaul implies minimal
possible cancellation of power from the composite message (and
thus at the receiver $B$) and the remaining power of the wireless
signal from $C$ at $B$ is largest possible. At relatively low
$R_C$, this effect increases the interval of values for $R_C$ that
are completely decodable at $B$ and that is why sending $W_{C2}$
over the backhaul gives higher achievable rate $S_{AB}(R_C)$.
However, when the rate $R_C$ is large and thus not completely
decodable at $B$, then $X_{C1}$ acts as a noise and such a high
remaining power harms the rate achievable for large $R_{C}$. On
the other hand, when $W_{C1}$ is sent over the backbone, the
uncancelled part of the composite message has minimum possible
power, which is desirable when that portion of the signal sent by
$C$ is undecodable and has to be treated as noise.

\section{Numerical Examples\label{sec: numerical}}

In this section we provide some numerical examples for the
performance of the proposed confidential transmission schemes when
all the wireless channels are deterministic (unfaded), as assumed
in the previous sections. We will use the following acronyms:
\emph{EQ} for Elementary Quantization, \emph{WZ} for Wyner--Ziv
quantization, and \emph{SUP} for transmission based on
superposition coding. In the cases $R_C \leq C_L$, the message
from $C$ is completely transferred via the backhaul, such that all
the schemes behave identically.

\begin{figure}[ptb]
\centering
\includegraphics[width=8.3 cm]{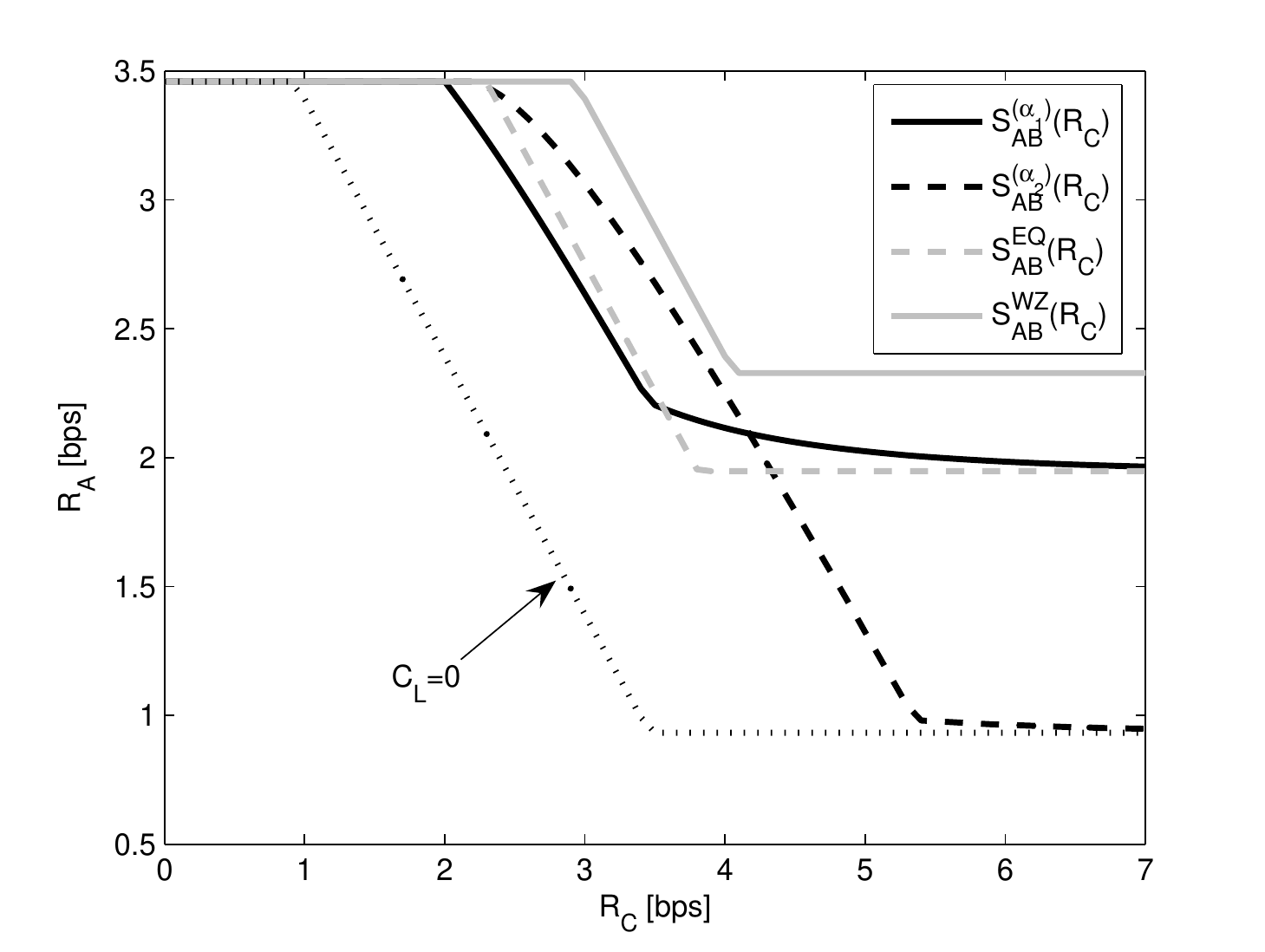} \caption{Maximum
achievable rates from $A$ to $B$ (without confidentiality constraints)
$S_{AB}^{(\alpha_{1})}(R_{C}),$ $S_{AB}^{(\alpha_{2})}(R_{C})$, $S_{AB}%
^{EQ}(R_{C})$ and $S_{AB}^{WZ}(R_{C})$ with $\gamma_{AB}=\gamma_{CB}=10$ [dB]
and $C_{L}=2$ [bps]. As a reference, the function $S_{AB}(R_{C})$ with
$C_{L}=0$ is also shown.}%
\label{fig:S_functions}%
\end{figure}

We start by considering the maximum achievable rates with no confidentiality
constrains from terminal $A$ to BS $B$, namely $S_{AB}^{EQ}(R_{C})$
(\ref{eq:Sq}) (\ref{gamma Q}); $S_{WZ}^{EQ}(R_{C})$ (\ref{eq:Sq})
(\ref{eq:WZgammaq}); $S_{AB}^{(\alpha_{1})}(R_{C})$ (\ref{eq: S_AB1}) and
$S_{AB}^{(\alpha_{2})}(R_{C})$ (\ref{eq: S_AB2}). Fig.~\ref{fig:S_functions}
depicts such rates versus the downlink rate $R_{C}$. For the chosen parameters
($\gamma_{AB}=\gamma_{CB}=10$ [dB] and $C_{L}=2$ [bps]), WZ is to be preferred
for any value of the downlink rate $R_{C}$. Moreover, by appropriately
selecting which message is sent over the backbone ($W_{C1}$ or$\ W_{C2}$),
that is choosing between $S_{AB}^{(\alpha_{1})}(R_{C})$ and $S_{AB}%
^{(\alpha_{2})}(R_{C}),$ the SUP strategy outperforms the EQ for any $R_{C}$.
On this note, confirming the discussion of Sec. \ref{prop:quant}, we have that
for lower $R_{C}$ it is more convenient to send $W_{C2}$ over the backbone
($S_{AB}^{(\alpha_{2})}(R_{C})>S_{AB}^{(\alpha_{1})}(R_{C}))$ and viceversa
for larger $R_{C}.$ Finally, we remark that, as pointed out in Sec.
\ref{sec: achievable SUP}, if the rate $R_{C}$ is large enough, the EQ
strategy obtains a constant secrecy rate $R_{C}$, which coincides with the
asymptotic achievable for of the SUP strategy.

\begin{figure}[ptb]
\centering
\includegraphics[width=8.3 cm]{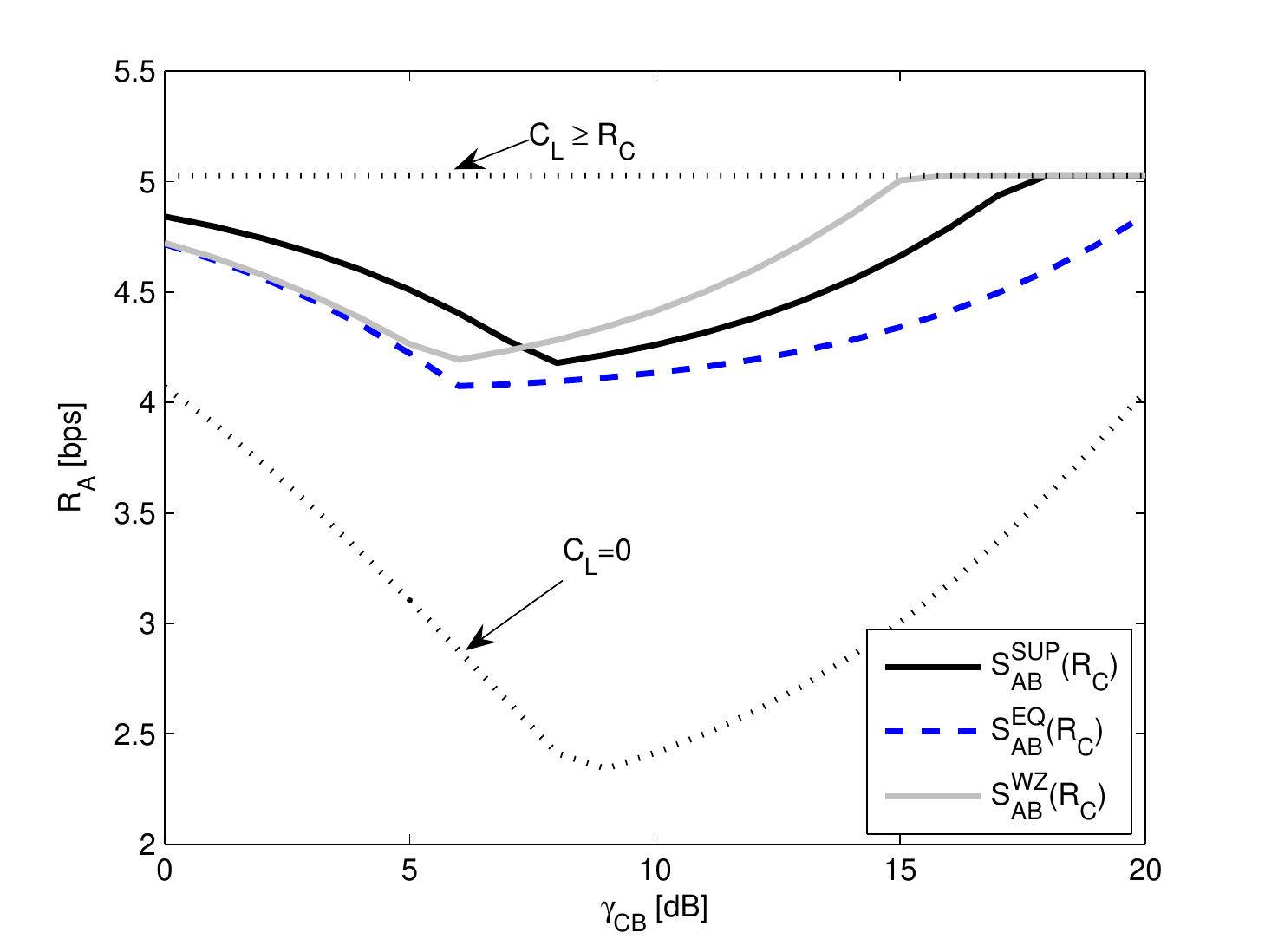} \caption{Maximum
achievable rates from $A$ to $B$ (without confidentiality constraints)
$S_{AB}^{(\alpha_{1})}(R_{C}),$ $S_{AB}^{(\alpha_{2})}(R_{C})$, $S_{AB}%
^{Q}(R_{C})$ and $S_{AB}^{WZ}(R_{C})$ with $\gamma_{AB}=15$ [dB], $C_{L}=2$
[bps] and $R_{C}=3$ [bps]. As a reference, we have plotted the line
$S_{AB}(R_{C})$ with $C_{L}=0$ and $S_{AB}(R_{C})=\mathcal{C}(\gamma_{AB})$
for $C_{L}\geq R_{C}$.}%
\label{fig:RA_vs_gCB_CL2_RC3_gAB15}%
\end{figure}

Fig.~\ref{fig:RA_vs_gCB_CL2_RC3_gAB15} shows the achievable rate without
confidentiality constraints (as Fig. \ref{fig:S_functions}) versus the SNR
between the BSs $\gamma_{CB}$. Here it can be seen that, for low values of
$\gamma_{CB}$ the SUP strategy outperforms the WZ strategy. The U-shape of all
the curves versus $\gamma_{CB}$ can be explained similarly to the arguments
used to study interference channels with with weak and strong interference.
Consider for instance the case $C_{L}=0$. For low $\gamma_{CB}$, BS $B$ cannot
decode $R_{C}$, but the wireless interference from $C$ at $B$ is weak, which
makes the achievable rate $A-B$ high. As $\gamma_{CB}$ increases, but still
not sufficiently as to make rate $R_{C}$ decodable at $B$, the achievable rate
on link $A-B$ drops. However, for strong interference $\gamma_{CB}$, BS $B$
can decode $R_{C}$ and then subtract it, thus causing low (if any) penalty to
the rate from $A$ to $B$.

\begin{figure}[ptb]
\centering
\includegraphics[width=8.3 cm]{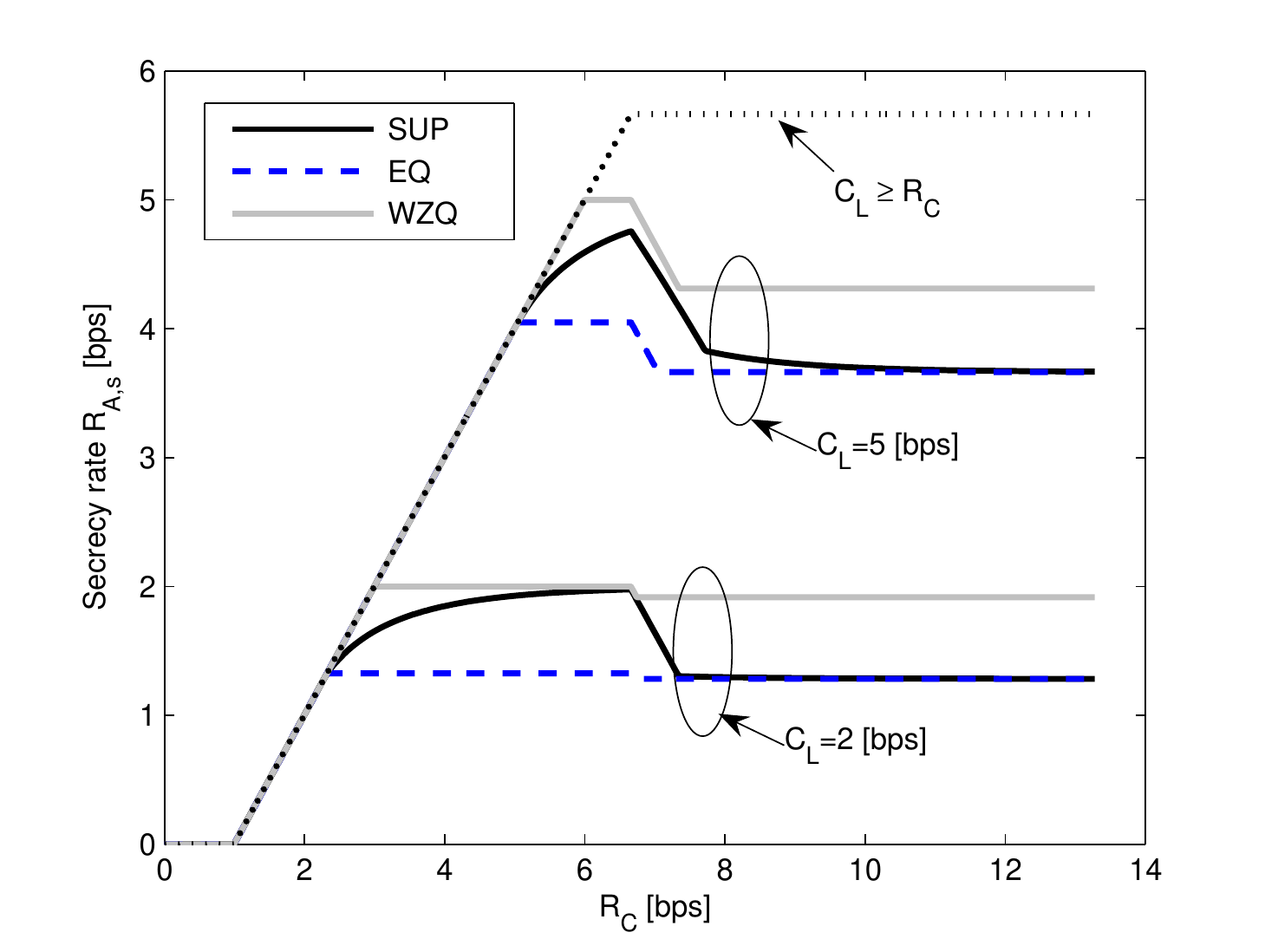}
\caption{Achievable secrecy rates by superposition, elementary quantization
and Wyner-Ziv quantization for different $C_{L}$ versus $R_{C}$ ($\gamma
_{AB}=\gamma_{CB}=\gamma_{AE}=\gamma_{CE}=20$ [dB]).}%
\label{fig:secrecy_rate}%
\end{figure}

Finally, Fig.~\ref{fig:secrecy_rate} depicts the derived secrecy
rates for different values of $C_{L}$ versus the downlink rate
$R_{C}$ for $\gamma _{AB}=\gamma_{CB}=\gamma_{AE}=\gamma_{CE}=20$
[dB]. Note that with such a choice of SNRs the Noise--Forwarding
strategy~\cite{ElGamal-RelayEavesdrop} $(C_{L}=0)$ offers a zero
secrecy rate, which implies that in this case the presence of the
backbone offers markedly improved secrecy. Moreover, for small
downlink rates $R_{C}$, all proposed strategies have the same
achievable secrecy rate as in the case of large backbone capacity
studied in Sec. \ref{sec: large backbone}\ ($C_{L}>R_{C}$) up to a
certain value of $R_{C}$, which is the largest for the WZ
strategy. Finally, as pointed out above, WZ\ offers substantial
gains with respect to EQ, and, given the large value of
$\gamma_{CB}$ in this example, also with respect to SUP (where SUP
and EQ have the same performance for large $R_{C}).$

%As a final remark, it should be noted that the price to be paid
%for the better secrecy performance of the superposition strategy
%is that $C$ should explicitly account for the rate of $A$ to $B$
%when transmitting to $D$ and create the codebooks accordingly. For
%the quantization strategy, the transmission $C-D$ is oblivious
%with respect to the secrecy objective, which could be an asset for
%practical system implementation.

\section{Extension to Fading Channels}
\label{sec:ExtensionFading}

In this section, we turn the attention to fading channels and reconsider the
performance of the proposed transmission strategies under different
assumptions regarding the channel state information available at different nodes.

\subsection{Scenario and Performance Measures}

The inter-BS link $C-B$ is considered to be a line--of--sight and thus does
not experience fading, i.~e., $\gamma_{CB}$ is constant, while the other links
are faded. We assume that a fading link $h_{UV}$ features Rayleigh fading,
such that the SNR of the link $\gamma_{UV}$ is independently and exponentially
distributed with average value $\bar{\gamma}_{UV}$. Furthermore, we consider
block fading, such that a fading channel stays constant for a sufficient
number of symbols $n$, where for coding purposes $n$ can be assumed to be
infinity. It is noted that the assumption regarding the inter-BS link
$\gamma_{CB}$ is a reasonable if, e.g., the BSs are sufficiently elevated with
respect to the rest of the network. As far as channel state information is
concerned, terminal $A$ is assumed to know the channel gains $\gamma_{AB}$
(and the constant $\gamma_{CB}),$ beside the downlink rate $R_{C},$ so that it
can calculate (and transmit at) the maximum instantaneous achievable rate
$S_{AB}(R_{C})$ in (\ref{eq:ExampleS_U1V}). Other assumptions will be
differently specified below for two scenarios, one in which we measure the
outage probability and the other in which we assess the scheduling performance.

\subsubsection{No Channel State Information about \textrm{E}: Outage
Probability}

This scenario relies on the realistic assumption that the instantaneous fading
channel to the eavesdropper $\gamma_{AE}$ and $\gamma_{CE}$ are not known to
terminal $A$ and BS $C$. In such a case, no non-zero rate is achievable with
perfect secrecy, and therefore we one has to resort to the concept of outage
probability~\cite{barros, letter}. In particular, given a target secrecy rate
$R_{A,s}$, the outage probability is defined as the probability that such
$R_{A,s}$ is not achievable for the given transmission technique. It is noted
that, for each fading realization, the value of $R_{C}$ is selected is here
selected as (\ref{eq:SUPtotalrate}), which requires BS C to know the
instantaneous downlink channel $\gamma_{CD}-D$.

\subsubsection{Full\ Channel State Information: Scheduling Performance}

In this second scenario, we assume full channel state information about all
the fading channels at both terminal $A$ and BS $C$ know. Given the full
channel state information, it is relevant here to generalize the model to
include $M_{u}$ uplink users $A_{1},A_{2},\ldots A_{M_{u}}$ that have data to
transmit to $B$ and $M_{d}$ downlink users $D_{1},D_{2},\ldots D_{M_{d}}$
potentially receiving from BS $C$. The goal is to analyze the impact of
different scheduling and transmission strategies on the performance of the
network at hand over fading channels. As throughout the paper, of particular
interest is the impact of design choices on the trade-off between the downlink
($R_{C})$ and the uplink secrecy rate ($R_{A,s})$.

Regarding uplink scheduling, we assume that the uplink user $A_{i^{\ast}}$ is
selected so as to maximize the uplink rate:
\begin{equation}
i^{\ast}=\max_{i}\gamma_{A_{i}B} \label{eq:UplinkSelection}%
\end{equation}
More interesting is the scheduling of the downlink transmissions from $C$, for
which we define two different types of schedulers:

\begin{itemize}
\item \emph{Max$R_{C}$} scheduler: In this case the scheduled user $D_{j\ast}$
is selected so as to maximize the downlink rate:
\begin{equation}
j^{\ast}=\max_{j}\gamma_{CD_{j}} \label{eq:MaxCSI}%
\end{equation}

\item \emph{MaxSec} scheduler: In this case the selection of the user
$D_{j^{\ast}}$ is done so as to maximize the uplink secrecy rate $R_{A,s}$.
Accordingly, the selection of $D_{j^{\ast}}$ depends on which method is used
by $C$ to communicate over the backbone. If WZ quantization is used, the
scheduler is denoted $MaxSec_{WZ}$ and we have:
\begin{equation}
j^{\ast,WZ}=\max_{j}R_{A,s}^{WZ}(\log_{2}(1+\gamma_{CD_{j}})),
\label{eq:WynerZivSElectD}%
\end{equation}
while if superposition is used, the scheduler is denoted $MaxSec_{SUP}$ and:
\begin{equation}
j^{\ast,SUP}=\max_{j}R_{A,s}^{sup}(\log_{2}(1+\gamma_{CD_{j}})).
\label{eq:WynerZivSElectD}%
\end{equation}

\end{itemize}

Note that, in general $j^{\ast,WZ}\neq j^{\ast,SUP}$. Moreover, notice that
only the WZ strategy has been considered among the quantization schemes to
simplify the discussion and given the superior performance with respect to EQ.
Performance evaluation is then carried out by calculating the average secrecy
rate $\bar{R}_{A,s}(\mathcal{S})$ and the average downlink rate $\bar{R}%
_{C}(\mathcal{S})$, where the average is taken with respect to the fading
channels ($\gamma_{A_{i^{\ast}}B},$ $\gamma_{A_{i^{\ast}}E},$ $\gamma_{CE},$
$\gamma_{CD_{j^{\ast}}}$) given the scheduler $\mathcal{S}\in\{MaxR_{C}%
,MaxSec_{sup},MaxSec_{WZ}\}$.

\subsection{Numerical Results}

We now present some numerical results for the two considered scenarios.

\subsubsection{Outage Probability}

\begin{figure}[ptb]
\centering
\includegraphics[width=8.3 cm]{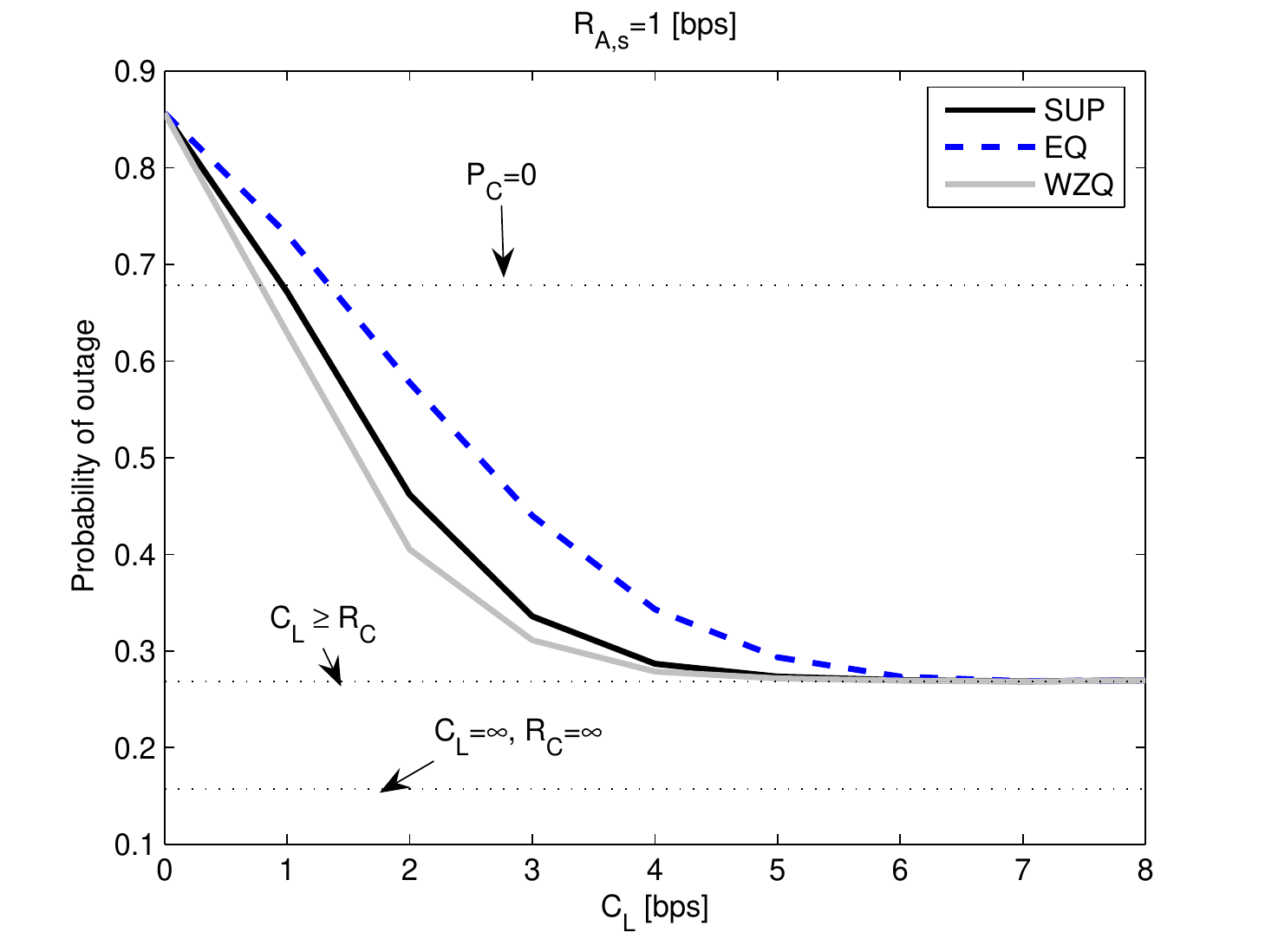}
\caption{Secrecy outage probability vs. the backhaul capacity $C_{L}$. The
average values of the fading links are $\bar{\gamma}_{AB}=\bar{\gamma}%
_{AE}=\bar{\gamma}_{CE}=\bar{\gamma}_{CD}=15$ [dB]. The constant value
$\gamma_{CB}$ is 15 [dB]. The line $P_{C}=0$ refers to the case when no
cooperative interference from $C$ takes place. The target secrecy rate is
$R_{A,s}=1$ [bps].}%
\label{fig:fading_Pout_vs_CL}%
\end{figure}

Fig.~\ref{fig:fading_Pout_vs_CL} depicts the outage probability as a function
of the backbone capacity $C_{L}$ for $\bar{\gamma}_{AB}=\bar{\gamma}_{AE}%
=\bar{\gamma}_{CE}=\bar{\gamma}_{CD}=15$ [dB], $\gamma_{CB}=15$ [dB],
$R_{A,s}=1$ [bps]. We recall that the value $R_{C}$ is selected according to
the instantaneous SNR $\gamma_{CD}$ as (\ref{eq:SUPtotalrate}). The line
$C_{L}\geq R_{C}$ is obtained by assuming that $C_{L}$ is large enough to can
accommodate any rate $R_{C}$ (strictly speaking, $C_{L}\rightarrow\infty)$. It
can be seen that, as $C_{L}$ increases, the outage probability of all the
strategies approaches this asymptotic performance, as it becomes highly
probable that the given $C_{L}$ can accommodate the rate $\mathcal{C}%
(\gamma_{CD})$. The lower bound on the outage probability is obtained by
assuming that $C$ sends pure Gaussian noise (which is the worst jamming
signal, see, e.g., \cite{Diggavi}), that is perfectly transferred through the
backbone ($C_{L}=\infty,R_{C}=\infty$). Another reference performance is set
by the case $P_{C}=0,$ where no downlink transmission takes place. For low
values of $C_{L}$, the proposed schemes can actually be outperformed by such
solution. This is because, for low $C_{L},$ the downlink transmission impairs
not only reception at the eavesdropper $E$, but also at the BS $B$.

\begin{figure}[ptb]
\centering
\includegraphics[width=8.3 cm]{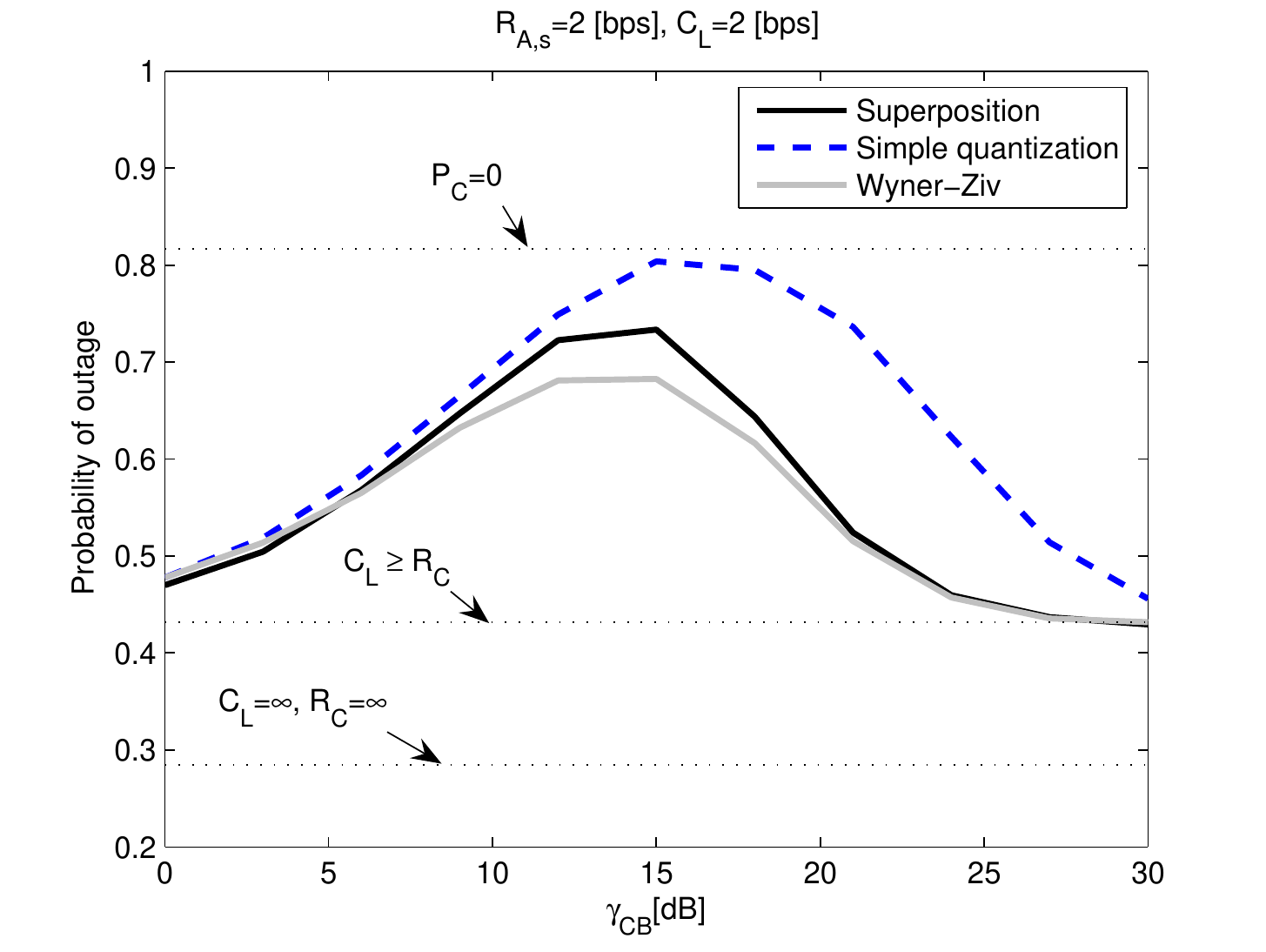}
\caption{Secrecy outage probability vs. the SNR $\gamma_{CB}$. The average
values of the fading links are $\bar{\gamma}_{AB}=\bar{\gamma}_{AE}%
=\bar{\gamma}_{CE}=\bar{\gamma}_{CD}=15$ [dB]. The constant value $\gamma
_{CB}$ is 15 [dB]. The line $P_{C}=0$ refers to the case when no cooperative
interference from $C$ takes place. The target secrecy rate is $R_{A,s}=2$
[bps]. The backhaul has $C_{L}=2$ [bps], except for the reference line with
$C_{L}=\infty$.}%
\label{fig:fading_Pout_vs_gCB}%
\end{figure}

Fig.~\ref{fig:fading_Pout_vs_gCB} depicts the outage probability as a function
of the inter-BS SNR $\gamma_{CB}$ for $\bar{\gamma}_{AB}=\bar{\gamma}%
_{AE}=\bar{\gamma}_{CE}=\bar{\gamma}_{CD}=15$ [dB], $R_{A,s}=1$ [bps] and
$C_{L}=2$ [bps]. The U-shape of the curves for the proposed strategies can be
explained by resorting to similar arguments as for
Fig.~\ref{fig:RA_vs_gCB_CL2_RC3_gAB15} (see Sec. \ref{sec: numerical}).
Following this remark, we note from Fig.~\ref{fig:fading_Pout_vs_gCB} that the
gain in terms of outage probability of all strategies with respect to the case
$C_{L}=0$ is most relevant in the regime of weak/strong interference from BS
$C$ (i. e. low/high $\gamma_{CB}$). In fact, it is in this regime that the
interference from BS\ $C$ to $B$ (due to the realizations where $R_{C}>C_{L}%
)$\ has the least impact on the performance of the link $A-B$.

\subsubsection{Scheduling performance}

\begin{figure}[ptb]
\centering
\includegraphics[width=8.3 cm]{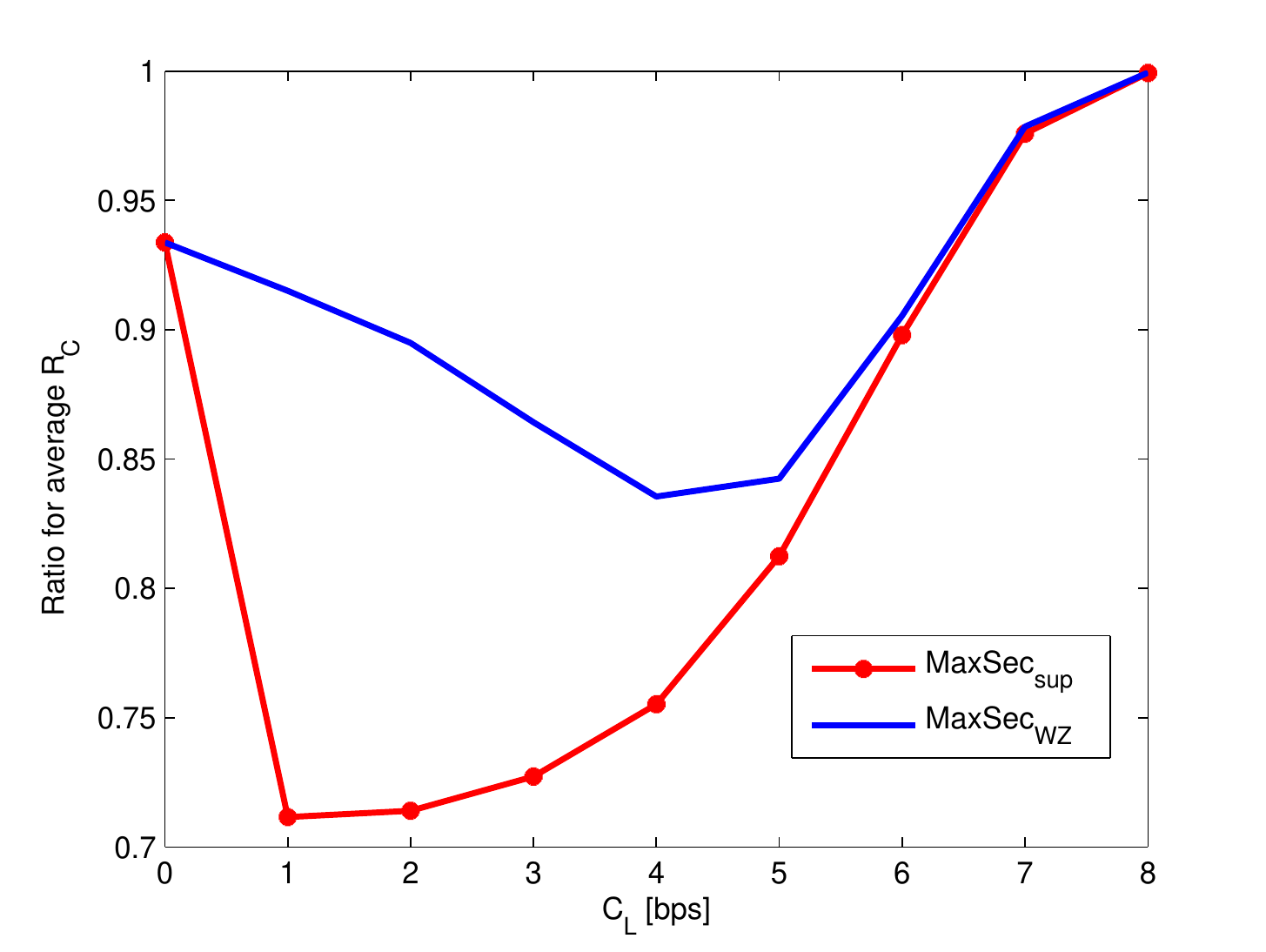}\caption{Ratios
$\frac{\bar{R}_{C}(MaxSec_{SUP})}{\bar{R}_{C}(MaxR_{C})}$ and $\frac{\bar
{R}_{C}(MaxSec_{WZ})}{\bar{R}_{C}(MaxR_{C})}$. The parameters are $\bar
{\gamma}_{AB}=\bar{\gamma}_{AE}=\bar{\gamma}_{CD}=\bar{\gamma}_{CE}=15$ [dB].
The constant SNR is $\gamma_{CB}=5$ [dB].}%
\label{fig:sched_1_ratioRC_15_15_15_15_5}%
\end{figure}

Turning to the average rates that can be achieved in the scenario of full
channel state information, Fig.~\ref{fig:sched_1_ratioRC_15_15_15_15_5}
considered the downlink rates in terms of the ratios $\frac{\bar{R}%
_{C}(MaxSec_{sup})}{\bar{R}_{C}(MaxR_{C})}$ and $\frac{\bar{R}_{C}%
(MaxSec_{WZ})}{\bar{R}_{C}(MaxR_{C})}$ for
$\bar{\gamma}_{AB}=\bar{\gamma
}_{AE}=\bar{\gamma}_{CD}=\bar{\gamma}_{CE}=15,$ $\gamma_{CB}=5$
[dB]. These ratio demonstrate which fraction of the maximal
average downlink throughput is achieved if the scheduler at $C$
aims to maximize the secrecy of the transmission $A-B$.
Equivalently, the complement to one of such ratios measure the
fractional rate loss due to the requirement of maximizing the
secrecy of the transmission $A-B.$ The results show that at high
$C_{L}$, maximum secrecy is coherent with maximal rate $R_{C}$.
However, from Fig.~\ref{fig:sched_1_ratioRC_15_15_15_15_5} it is
seen that for lower $C_{L}$ maximal security is not always
achieved by maximizing $R_{C}$, which is in accordance with the
observations from Fig.~\ref{fig:secrecy_rate}. Regarding the SUP
strategy, there is one degenerative effect, which can be explained
by observing the SUP curve on Fig.~\ref{fig:secrecy_rate}. It can
be seen (on the figure not discernible for $C_L=2$) that for large
$R_C$, the secrecy rate of the SUP scheme slowly decreases towards
the asymptotic value (achieved for $R_C \rightarrow \infty$),
while for the quantization schemes there are finite values of
$R_C$ after which the secrecy rate becomes constant. Hence, the
scheduler that maximizes the secrecy tends to select lower rates
$R_C$ when SUP is applied. Nevertheless, when $C_L=0$, both SUP
and WZ operate in identical way.

%It can be
%seen that when $C_L>0$, after the rightmost knee--point
%(non-differentiable point), the secrecy rate decreases
%monotonically as a function of $R_C$. Therefore, the
%$MaxSec_{SUP}$ scheduler always schedules user with $R_C$ lower
%than the maximum, in order to approach better the knee point.
%However, as $C_L$ approaches 0, the gain in the secrecy rate is
%marginal, although the selected ``optimal'' $R_C$ can be
%significantly smaller than the maximal $R_C$ across the users.
%But, when $C_L=0$ (not plotted on Fig.~\ref{fig:secrecy_rate}),
%there is a value of $R_C$ after which $R_{A,s}$ stays constant,
%analogously to the curves for WZ or EQ on
%Fig.~\ref{fig:secrecy_rate}. In such a case...ZAVRSI GO OVA!

%It is worth noting that the loss
%in the rate $R_C$ due to secrecy maximization depends
%significantly on the wireless link between the access points i.~e.
%$\gamma_{CB}$.

\begin{figure}[ptb]
\centering
\includegraphics[width=8.3 cm]{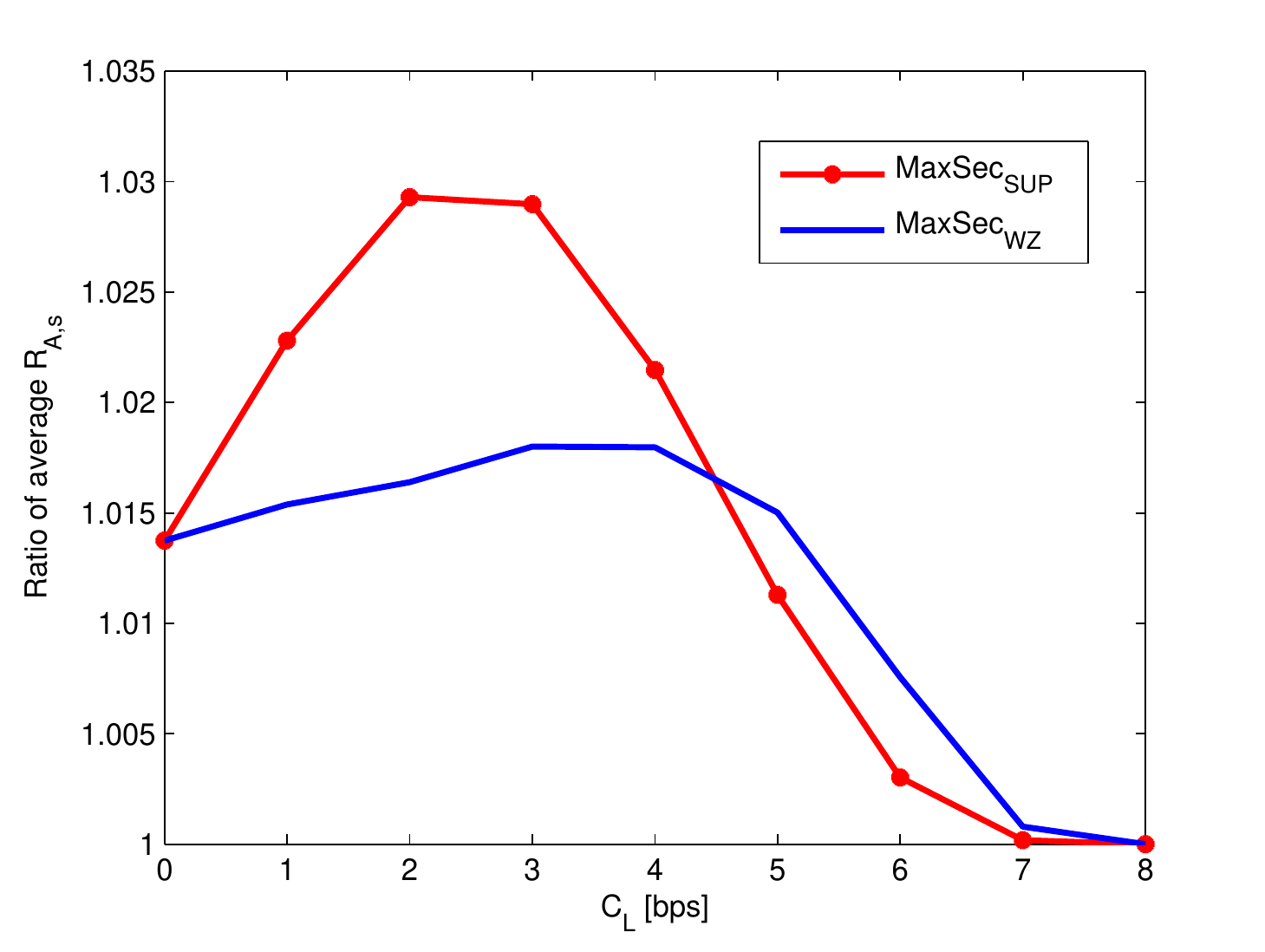}\caption{Ratios
$\frac{\bar{R}_{A,s}(MaxSec_{sup})}{\bar{R}_{A,s}(MaxR_{C})}$ and $\frac
{\bar{R}_{A,s}(MaxSec_{WZ})}{\bar{R}_{A,s}(MaxR_{C})}$. The parameters are
$\bar{\gamma}_{AB}=\bar{\gamma}_{AE}=\bar{\gamma}_{CD}=\bar{\gamma}_{CE}=15$
[dB]. The constant SNR is $\gamma_{CB}=5$ [dB].}%
\label{fig:sched_1_ratioRAs_15_15_15_15_5}%
\end{figure}

\begin{figure}[ptb]
\centering
\includegraphics[width=8.3 cm]{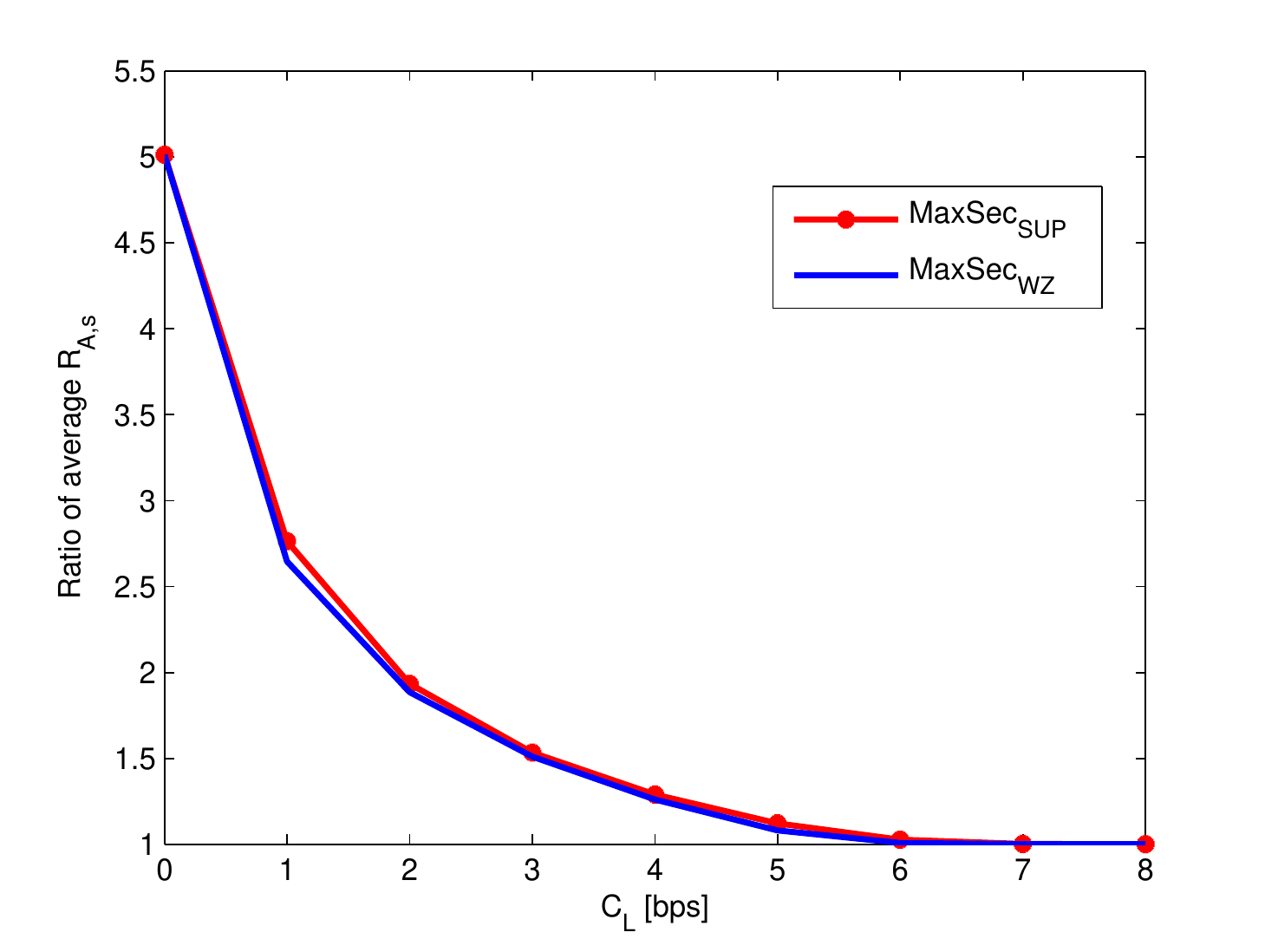}\caption{Ratios
$\frac{\bar{R}_{C}(MaxSec_{sup})}{\bar{R}_{C}(MaxR_{C})}$ and $\frac{\bar
{R}_{C}(MaxSec_{WZ})}{\bar{R}_{C}(MaxR_{C})}$. The parameters are $\bar
{\gamma}_{AB}=\bar{\gamma}_{AE}=\bar{\gamma}_{CD}=\bar{\gamma}_{CE}=15$ [dB].
The constant SNR is $\gamma_{CB}=20$ [dB].}%
\label{fig:sched_3_ratioRAs_15_15_15_15_20}%
\end{figure}

The average secrecy rates from $A$ to $B$ are then shown in
Fig.~\ref{fig:sched_1_ratioRAs_15_15_15_15_5}%
-\ref{fig:sched_3_ratioRAs_15_15_15_15_20} in terms of the ratios $\frac
{\bar{R}_{A_{s}}(MaxSec_{sup})}{\bar{R}_{A,s}(MaxR_{C})}$ and $\frac{\bar
{R}_{A,s}(MaxSec_{WZ})}{\bar{R}_{A,s}(MaxR_{C})}$. Thus, the figures shows,
for each transmission method (WZ or SUP), how much the secrecy rate is
improved if the scheduler at $C$ determines the downlink user (and the
corresponding rate $R_{C})$ in order to maximize the instantaneous rate
$R_{A,s}$ rather than the downlink rate $R_{C}$. It can be seen that for a
weak link $C-B$ (low $\gamma_{CB}$), as on
Fig.~\ref{fig:sched_1_ratioRAs_15_15_15_15_5} (where $\gamma_{CB}=5$ [dB]),
the gain in the secrecy rate for the $MaxSec$ schedulers is insignificant,
which means that application of opportunistic scheduler $MaxR_{C}$ at $C$ will
be also good for the security of the link $A-B$. Conversely, for a strong link
$C-B$, the results on Fig.~\ref{fig:sched_1_ratioRAs_15_15_15_15_5}%
-\ref{fig:sched_3_ratioRAs_15_15_15_15_20} show that the secrecy of the link
$A-B$ can be boosted by selecting appropriate non-maximal $R_{C}$ and in this
case the opportunistic downlink scheduler at $C$ is not compatible with the
secrecy requirements.

\section{Conclusions}

Optimized scheduling and multi-cell BS cooperation are becoming increasingly
standard features of current and future wireless infrastructure (cellular)
networks. This work has advanced the notion that such technologies can play an
important role in ensuring confidentiality (security) of wireless
transmissions. From the analysis of several transmission strategies under
different assumptions regarding propagation channels and corresponding channel
state information, a number of conclusions have been drawn. In particular, a
technique based on Wyner-Ziv compression over the backbone link connecting the
BSs has proved to be the most promising, with the added benefit of requiring
no modifications on the uplink/ downlink transmissions of a conventional
cellular systems. When complexity of Wyner-Ziv encoding is an issue, one could
resort to simpler quantization schemes with a performance loss that depends on
the network topology. Or else, if willing to modify the downlink transmission/
reception strategy for the sake of ensuring uplink confidentiality, one could
opt for channel coding (rather than source coding) based techniques which
perform close (or even better than) Wyner-Ziv under some circumstances.

There are several interesting extensions of this work. In this
paper we have shown some achievable rates, but it is important to
know what is the true secrecy capacity of the introduced method or
at least to derive some tight upper bounds. Furthermore, the
considered scenario can be extended to multiple cooperating base
stations, which raises the question how to organize the
transmission/receive schedule for the access points in order to
maximize the secrecy effect, while not degrading the throughput.
Finally, the study can be extended to consider colluding
eavesdroppers, which attempt to jointly decode the desired signal
and the interference from the downlink transmissions.

\appendices

\section{Appendix-A: Proof of Proposition \ref{prop:quant}}

The equivalent signal seen at BS $B$ over both the wireless and wired channels
in a given time instant $i$ can be written as a vector MAC channel:
\begin{equation}
\tilde{Y}_{B,i}=\left[
\begin{array}
[c]{c}%
Y_{B,i}\\
\hat{X}_{C,i}%
\end{array}
\right]  =\left[
\begin{array}
[c]{cc}%
h_{AB} & h_{CB}\\
0 & 1
\end{array}
\right]  \left[
\begin{array}
[c]{c}%
X_{A,i}\\
X_{C,i}%
\end{array}
\right]  +\left[
\begin{array}
[c]{c}%
N_{i}\\
Q_{i}%
\end{array}
\right]  . \label{eq: Y tilde}%
\end{equation}
Let $S_{AB}^{EQ}(R_{C})$ denote the maximum achievable rates from $A$ to $B$
for a given transmission rate $R_{C}$ when the quantization strategy is used
(recall (\ref{eq:ExampleS_U1V})). In order to determine $S_{AB}^{EQ}(R_{C})$,
we have to examine the achievable region for the vector MAC channel with
output (\ref{eq: Y tilde}):
\begin{subequations}
\label{vector MAC}%
\begin{align}
R_{AB}  &  <I(X_{A};\tilde{Y}_{B}|X_{C})\label{eq:EQvectorMAC}\\
R_{CB}  &  <I(X_{C};\tilde{Y}_{B}|X_{A})\\
R_{AB}+R_{CB}  &  <I(X_{A},X_{C};\tilde{Y}_{B})
\end{align}
where we have dropped the index $i$ for simplicity) and $X_{A}$ represents
normally--distributed complex signal transmitted by $A$. The mutual
informations in (\ref{Sec:Wc2}) can be determined as follows:
\end{subequations}
\begin{equation}
I(X_{A};\tilde{Y}_{B}|X_{C})=I(X_{A},Y_{B}|X_{C})=\mathcal{C}(\gamma_{AB})
\label{eq:EQvectorMAC_1}%
\end{equation}
since $\hat{X}_{C}$ is conditionally independent of $X_{A}$ when $X_{C}$ is
given. The second bound leads to:
\begin{equation}
I(X_{C};\tilde{Y}_{B}|X_{A})=\mathcal{C}(\gamma_{CB}+\gamma_{Q})
\label{eq:EQvectorMAC_2}%
\end{equation}
while the third condition is:
\begin{align}
I(X_{A},X_{C};\tilde{Y}_{B})  &  =\label{eq:C sum}\\
&  =I(X_{C};\tilde{Y}_{B})+I(X_{A};\tilde{Y}_{B}|X_{C})=\\
&  \overset{(a)}{=}\mathcal{C}\left(  \frac{\gamma_{CB}}{1+\gamma_{AB}}%
+\gamma_{Q}\right)  +I(X_{A};Y_{B}|X_{C})\\
&  =\mathcal{C}\left(  \frac{\gamma_{CB}}{1+\gamma_{AB}}+\gamma_{Q}\right)
+\mathcal{C}(\gamma_{AB})\\
&  =\log_{2}\left(  2^{C_{L}}(1+\gamma_{AB})+\gamma_{CB}\right)
\end{align}
where (a) follows again from $X_{A}$ being conditionally independent of
$\hat{X}_{C}$ for given $X_{C}$. The rate $S_{AB}^{EQ}(R_{C})$ in
(\ref{prop:quant}) then easily follows.

\section{Appendix-B: Proof of (\ref{eq:WZderivation})}

Using the same model for the vector MAC channel as in~(\ref{eq: Y tilde}), we
can write:
\begin{align}
C_{L}  &  =I(X_{C};\hat{X}_{C}|Y_{B})=I(X_{C};\hat{X}_{C},Y_{B})-I(X_{C}%
;Y_{B})=\nonumber\label{eq:WZderivation}\\
&  =\mathcal{C}\left(  \frac{\gamma_{CB}}{1+\gamma_{AB}}+\gamma_{Q}\right)
-\mathcal{C}\left(  \frac{\gamma_{CB}}{1+\gamma_{AB}}\right)  =\nonumber\\
&  =\log_{2}\left(  1+\frac{\gamma_{Q}}{1+\frac{\gamma_{CB}}{1+\gamma_{AB}}%
}\right)
\end{align}
and thus (\ref{eq:WZderivation}) easily follows.

\section{Appendix-C: Proof of Lemma \ref{prop: SUP1}}

Similarly to Appendix-A, we need to determine the maximal achievable rate
$S_{AB}^{(\alpha_{2})}(R_{C}).$ We start from the MAC capacity region obtained
after the cancellation of $W_{C2}$:
\begin{align}
R_{A}  &  <\mathcal{C}(\gamma_{AB})\label{eq:RateRA_afterXC1cancel}\\
R_{C1}=R_{C}-C_{L}  &  <\mathcal{C}(\alpha_{2}\gamma_{CB}%
)\label{eq:RateRC2_afterXC1cancel}\\
R_{A}+R_{C}-C_{L}  &  <\mathcal{C}(\gamma_{AB}+\alpha_{2}\gamma_{CB}),
\label{eq:RateRARC2_afterXC1cancel}%
\end{align}
Depending on the value of $R_{C},$ we have two cases to consider, namely, when
$R_{C1}$ is decodable at $B$ ($R_{C}-C_{L}<\mathcal{C}(\alpha_{2}\gamma_{CB}%
)$) and when $R_{C1}$ is not ($R_{C}-C_{L}\geq\mathcal{C}(\alpha_{2}%
\gamma_{CB})$). Analysis of these two cases easily leads to (\ref{eq: S_AB2}).


\begin{thebibliography}{99}                                                                                               %


\bibitem {wiretap}A. D. Wyner, \textquotedblleft The wire-tap
channel,\textquotedblright\ \textit{Bell System Technical Journal}, vol. 54,
no. 8, pp. 1355--1387, 1975.

\bibitem {wiretapgaussian}S. K. Leung-Yan-Cheong and M. E. Hellman,
\textquotedblleft The gaussian wiretap channel,\textquotedblright%
\ \textit{IEEE Trans. Inform. Theory}, vol. 24, pp. 451--456, Jul. 1978.

\bibitem {tekin}E. Tekin and A. Yener, "The general Gaussian multiple access
and two-way wire-tap channels: achievable rates and cooperative jamming,"
submitted [arXiv:cs/0610103v1].

\bibitem {Liang2}Y. Liang and H.\ V. Poor, "Multiple-access channels with
confidential messages," \textit{IEEE\ Trans. Inform. Theor}y, vol. 54, no. 3,
pp. 976-1002, March 2008.

\bibitem {Liang}Y. Liang, H. V. Poor, S. Shamai (Shitz), \textquotedblleft
Secrecy capacity region of parallel broadcast channels,\textquotedblright\ in
\textit{Proc. IEEE\ Information Theory and Applications Workshop} (ITW 2007),
pp. 245-250, Jan. 29- Feb. 2, 2007.

\bibitem {Liu}R. Liu and H. V. Poor, \textquotedblleft Multiple antenna secure
broadcast over wireless networks,\textquotedblright\ in \textit{Proc. of the
First International Workshop on Information Theory for Sensor Networks}, Santa
Fe, NM, June 18 - 20, 2007.

\bibitem {Khisti}A. Khisti and G. Wornell, \textquotedblleft The MIMOME
channel,\textquotedblright\ in \textit{Proc. 45th Annual Allerton Conference
on Communication, Control, and Computing}, Monticello, Illinois, Sept. 26-28, 2007.

\bibitem {Oggier}F. Oggier, B. Hassibi. \textquotedblleft The Secrecy Capacity
of the 2x2 MIMO Wiretap Channel,\textquotedblright\ in \textit{Proc. 45th
Annual Allerton Conference on Communication, Control, and Computing,
}Monticello, Illinois, Sept. 26-28, 2007.

\bibitem {kramer}G. Kramer, I. Maric and R. D. Yates, \textit{Cooperative
Communications},\ Foundations and Trends in Networking (FnT), Now Publishers,
Jun. 2007.

\bibitem {ElGamal-RelayEavesdrop}L. Lai and H. El Gamal, "The
relay--eavesdropper channel: Cooperation for secrecy," submitted [arXiv:cs/0612044v1]

\bibitem {jwcc}S. Shamai (Shitz), O. Somekh, O. Simeone, A. Sanderovich, B.M.
Zaidel and H. V. Poor, \textquotedblleft Cooperative multi-cell networks:
impact of limited-capacity backbone and inter-users links,\textquotedblright%
\ in \textit{Proc. Joint Workshop on Coding and Communications},
D\"{u}rnstein, Austria, October 14 - 16, 2007.

\bibitem {somekh}O. Somekh, O. Simeone, Y. Bar-Ness, A. Haimovich, U.
Spagnolini and S. Shamai, \textquotedblleft An information theoretic view of
distributed antenna processing in cellular systems,\textquotedblright\ in
\textit{Distributed Antenna Systems}, Auerbach Publications, CRC Press,
2007.\textit{\ }

\bibitem {barros}J. Barros and M. R. D. Rodrigues, "Secrecy capacity of
wireless channels," in \textit{Proc. IEEE Int. Symp. Inform. Theory} (ISIT), 2006.

\bibitem {el gamal}P. K. Gopala, L. Lai, H. El Gamal, "On the secrecy capacity
of fading channels," submitted [arXiv:cs/0702112v1]

\bibitem {negi1}R. Negi and S. Goel, "Secret communication using artificial
noise," in \textit{Proc. IEEE Veh. Techn. Conference}, vol. 3, pp. 1906-1910,
Sept. 2005.

\bibitem {globecom}M. L. J\o rgensen, B. Yanakiev, G. E. Kirkelund, P.
Popovski, H. Yomo, T. Larsen, "Shout to secure: physical--layer wireless
security with known interference", in \textit{Proc. IEEE Globecom 2007}.

\bibitem {letter}O. Simeone and P. Popovski,\textquotedblleft Secure
communications via cooperative base stations,\textquotedblright\ to appear in
\textit{IEEE Commun. Letters}.

\bibitem {gallager}R. Gallager. \emph{Information Theory and Reliable
Communication.} John Wiley \& Sons, Inc., 1968.

\bibitem {cover}T.Cover and J. Thomas. \emph{Elements of Information Theory.}
Wiley-Interscience; 2nd edition, 2006.

\bibitem {Diggavi}S. N. Diggavi and T. M. Cover, \textquotedblleft The worst
additive noise under a covariance constraint,\textquotedblright%
\ \textit{IEEE\ Trans. Inform. Theory}, vol. 47, no. 7, pp. 3072-3081, Nov. 2001.

\bibitem {Wyner-Ziv}A. Wyner and J. Ziv, \textquotedblleft The rate-distortion
function for source coding with side information at the
decoder,\textquotedblright\textit{\ IEEE Trans. Inform. Theory}, vol. 22, no.
1, pp. 1-10, Jan 1976.
\end{thebibliography}
\end{document}